\documentclass[fleqn,usenatbib]{mnras}

\usepackage{newtxtext,newtxmath}

\usepackage[T1]{fontenc}
\usepackage{ae,aecompl}
\pdfoutput=1

\usepackage{graphicx}	
\usepackage{amsmath}	
\usepackage{amssymb}	
\usepackage{enumitem}

\title[Clustering of \ensuremath{\rm{H}\alpha} emitters]{The dependence of galaxy clustering on stellar mass, star-formation rate and redshift at $z=0.8-2.2$, with HiZELS}

\author[R.K. Cochrane et al.]{R. K. Cochrane,$^{1}$\thanks{E-mail: rcoch@roe.ac.uk}
P. N. Best,$^{1}$
D. Sobral,$^{2,3}$
I. Smail,$^{4}$
J. E. Geach$^{5}$
J. P. Stott,$^{2}$
\newauthor
D. A. Wake$^{6,7}$
\\
$^{1}$SUPA, Institute for Astronomy, Royal Observatory Edinburgh, EH9 3HJ, UK\\
$^{2}$Department of Physics, Lancaster University, Lancaster, LA1 4YB \\
$^{3}$Leiden Observatory, Leiden University, P.O. Box 9513, NL-2300 RA Leiden, The Netherlands \\
$^{4}$Centre for Extragalactic Astronomy, Department of Physics, Durham University, South Road, Durham DH1 3LE, UK  \\
$^{5}$Centre for Astrophysics Research, Science \& Technology Research Institute, University of Hertfordshire, Hatfield, AL10 9AB, UK \\
$^{6}$Department of Physics, University of North Carolina Asheville, One University Heights, Asheville, NC 28804, USA. \\
$^{7}$Department of Physical Sciences, The Open University, Milton Keynes MK7 6AA, UK \\
}

\date{Accepted 2017 December 18. Received 2017 December 11; in original form 2017 September 21}

\pubyear{2018}

\volume{{\rm in press}}

\begin{document}
\label{firstpage}
\pagerange{\pageref{firstpage}--\pageref{lastpage}}
\maketitle

\begin{abstract}\\
The deep, near-infrared narrow-band survey HiZELS has yielded robust samples of $\rm{H}\alpha$-emitting star-forming galaxies within narrow redshift slices at $z=0.8,\, 1.47$ and $2.23$. In this paper, we distinguish the stellar mass and star-formation rate ($\rm{SFR}$) dependence of the clustering of these galaxies. At high stellar masses ($M_{*}/M_{\odot}\gtrsim2\times10^{10}$), where HiZELS selects galaxies close to the so-called star-forming main sequence, the clustering strength is observed to increase strongly with stellar mass (in line with the results of previous studies of mass-selected galaxy samples) and also with SFR. These two dependencies are shown to hold independently. At lower stellar masses, however, where HiZELS probes high specific SFR galaxies, there is little or no dependence of the clustering strength on stellar mass, but the dependence on SFR remains: high-SFR low-mass galaxies are found in more massive dark matter haloes than their lower SFR counterparts. We argue that this is due to environmentally driven star formation in these systems. We apply the same selection criteria to the EAGLE cosmological hydrodynamical simulations. We find that, in EAGLE, the high-SFR low-mass galaxies are central galaxies in more massive dark matter haloes, in which the high SFRs are driven by a (halo-driven) increased gas content.
\end{abstract}


\section{Introduction}
A rich array of work reveals that key observable galaxy properties including stellar mass, colour, star-formation rate, and morphology correlate with galaxy environments \citep{Melorose1978,Dressler1980,Baldry2006,Peng2010,Koyama2013,Scoville2013a,Darvish2016}, with massive, red, quiescent spheroids residing in the densest environments. Studies of galaxy environments can help constrain galaxy formation and evolution processes \citep[e.g.][]{Peng2010}. Yet quantifying galaxy environments on a galaxy-by-galaxy basis can be difficult, particularly at high redshifts, because the accuracy of such measurements is highly dependent on the depth and uniformity of the observations and the quality of the redshifts \citep[e.g.][]{Cooper2005}. \\
\indent The two-point correlation function, which quantifies the clustering strength of a population of galaxies, provides a fairly robust technique for identifying the typical dark matter halo environments of galaxy populations. On large scales, the two-point correlation function is dominated by the linear `two-halo term', which depends on the clustering of galaxies within different dark matter haloes. The two-halo term essentially measures the galaxy bias, a measure of the difference between the spatial distribution of galaxies and that of the underlying dark matter distribution. On small scales, the non-linear `one-halo term', which quantifies the clustering of galaxies within the same dark matter halo, dominates. Given an understanding of the way in which haloes of different mass cluster (which is reasonably well understood from $N$-body simulations within the cosmological model, e.g. \citealt{Bond1991,Lacey1994,Jenkins2001}), the observed (projected or angular) two-point correlation function enables us to derive the halo occupation of samples of galaxies from their observed clustering. This technique is known as Halo Occupation Distribution (HOD; \citealt{Ma2000,Peacock2000,Berlind2002,Cooray2002,Kravtsov2003}) modelling. The HOD framework then provides typical host dark matter halo masses for galaxy samples. It is also possible to derive estimates of central and satellite galaxy fractions from the small-scale `one-halo term' \citep[e.g.][]{Zheng2005,Tinker2010b}.\\
\indent Galaxy clustering measures provide a statistical description for a population of galaxies rather than quantifying environments on a galaxy-by-galaxy basis. Strong trends in clustering strength have been observed with galaxy morphological type \citep{Davis1976}, colour \citep{Zehavi2005,Coil2008,Simon2009,Hartley2010,Zehavi2011}, star-formation rate \citep{Williams2009,Dolley2014,Wilkinson2016} and stellar mass \citep{Wake2011,Mccracken2015,Coupon2015,Hatfield2015}, with the more recent studies reaching back to $z\sim2-3$. A limited number of studies of Lyman break galaxies have probed even further, back to $z\sim6-7$ \citep[e.g.][]{Harikane2015a,Harikane2017,Hatfield2017}. The largest samples have permitted the splitting of galaxy populations by more than one observed property. For example, \cite{Norberg2002}, using low-redshift ($z<0.15$) data from the 2dF survey \citep{Cole2000}, found that both early- and late-type galaxies display higher $r_{0}$ values and therefore stronger clustering at brighter $B$-band absolute magnitudes ($M_{B}$). \cite{Coil2008} found broadly consistent results at $z\sim1$ using the DEEP2 galaxy redshift survey \citep{Newman2012}, also confirming that at fixed $M_{B}$, red galaxies are more strongly clustered than blue galaxies.\\
\indent Splitting by multiple variables in this manner is important for galaxy evolution studies. A natural consequence of the apparent tight ($\sim0.4\,\rm{dex}$ scatter) correlation between stellar mass and star-formation rate of star-forming galaxies \cite[the `main sequence', e.g.][]{Brinchmann2004,Daddi2007,Elbaz2007,Karim2011} is that fundamental trends in one of these properties manifest as trends in the other. Galaxies with star-formation rates below the main sequence can also complicate observed trends: the fraction of galaxies that are passive increases towards higher stellar masses \citep{Peng2010,Sobral2011}, and this can give rise to trends with stellar mass which might not exist for the star-forming population only (e.g. the bending of the main sequence, \citealt{Lee2015}). Therefore, in this work, we aim to investigate the dependence of galaxy clustering on galaxy stellar mass and star-formation rate separately. \\
\indent The High-Redshift(Z) Emission Line Survey (HiZELS, \citealt{Sobral2012}; see Section \ref{sec:sample_selection}) identifies galaxies via their emission lines, yielding reliably-selected samples of $\rm{H}\alpha$ emitters within narrow redshift slices back to $z=2.2$. $\rm{H}\alpha$ (rest-frame wavelength $6562.8\AA$) is the brightest of the hydrogen recombination lines, which trace the young massive stellar population. Given that $\rm{H}\alpha$ is sensitive to star formation on short time-scales ($\sim10^{7}\rm{yr}$) and is also well-calibrated and less strongly extincted by dust than ultraviolet light \citep{Garn2010a}, it is often used as a tracer of star-formation. The $\rm{H}\alpha$ line is red-shifted out of the optical and into the near-infrared at $z\sim0.5$, making it ideal for probing star-forming galaxies at high redshift using wide-field near-infrared ground-based telescopes \citep[e.g.][]{Moorwood2000,Geach2008,Koyama2010,Koyama2011,Koyama2013a,Lee2012}. The well-defined redshift distributions of the HiZELS samples of $\rm{H}\alpha$-selected star-forming galaxies are ideal for studies of galaxy clustering, and the large numbers of emitters allows for the study of the population divided into many subsamples. \\
\indent \cite{Sobral2010} presented the first study of $\rm{H}\alpha$ luminosity-binned HiZELS galaxies and found evidence of higher clustering strengths for the strongest emitters at $z=0.84$. \cite{Geach2008} and \cite{Geach2012} performed the first clustering studies of $L_{\rm{H}\alpha}$-selected galaxies at $z=2.23$, though the sample size was insufficient to split by luminosity. In our previous paper (\citealt{Cochrane2017}, hereafter referred to as C17), we confirmed that the trends found by \cite{Sobral2010} hold to higher redshifts, using larger HiZELS samples at $z=0.8$, $z=1.47$ and $z=2.23$. Transforming clustering strengths to dark matter halo masses using HOD modelling, we found that halo mass increases broadly linearly with $L_{\rm{H}\alpha}$ at all three redshifts. Scaling by the characteristic `break' of the $\rm{H}\alpha$ luminosity function, $L_{\rm{H}\alpha}^{*}$, transforms these relations to a single trend, revealing a broadly redshift-independent monotonic relationship between $L_{\rm{H}\alpha}/L_{\rm{H}\alpha}^{*}$ and halo mass (\citealt{Sobral2010}; see also \citealt{Khostovan2017} for similar relations with other line emitters). For all of our samples, $L_{\rm{H}\alpha}^{*}$ galaxies reside in dark matter haloes of mass $\sim10^{12}M_{\odot}$, the known peak of the stellar mass - halo mass relation \cite[e.g.][]{Behroozi2010}. We also found low satellite fractions ($\sim5\%$) for these samples. This suggested that the star-formation rates of central galaxies are being driven by the mass accretion rates of their dark matter haloes \citep[see also][for details of a stellar-halo accretion rate coevolution model that matches observational data well]{Rodriguez-Puebla2016}. \\
\indent \cite{Sobral2010} used the $K$-band luminosities of HiZELS galaxies as a proxy for their stellar mass, finding an increase in galaxy clustering with increasing $K$-band luminosity, though the trend was significantly shallower than was observed for $\rm{H}\alpha$ luminosities. Preliminary investigations in C17 involved splitting our larger sample of galaxies at $z=0.8$ into two bins by observed $K$-band magnitude. Intriguingly, we found that the strong, roughly linear relationship between $\log_{10}L_{\rm{H}\alpha}$ and $r_{0}$ held for our two samples, with any differences between the two $K$-band magnitude bins being much smaller than the trend with $\rm{H}\alpha$ luminosity. \cite{Khostovan2017} present consistent results in their study of $\rm{H}\beta + [OII]$ and $\rm{[OIII]}$ emitters from HiZELS: clustering strength increases more significantly with emission line strength than with galaxy stellar mass. \\
\indent In this paper, we extend our previous work to study the clustering of HiZELS star-forming galaxies as a function of both $\rm{H}\alpha$ luminosity and stellar mass in more detail. Rather than using $K$-band observed magnitude as a proxy for stellar mass, we use a full SED-fitting approach to estimate stellar masses. We then compare our observational results to the output of the state-of-the-art cosmological hydrodynamical simulation EAGLE \citep{Crain2015,McAlpine2015,Schaye2015}. The structure of this paper is as follows. In Section \ref{sec:sample_selection} we provide a brief overview of the HiZELS survey and discuss our stellar mass estimates in some depth. In Section \ref{sec:clustering_one} we review the clustering and HOD-fitting techniques presented in C17 that we adopt here. In Section \ref{sec:results} we present our results, and in Section \ref{sec:eagle_comparison} we compare these to the output of the EAGLE simulation. Conclusions are drawn in Section \ref{sec:conclusions}. \\
\indent We use an $H_{0} = 70\rm{kms}^{-1}\rm{Mpc}^{-1}$, $\Omega_{M} = 0.3$ and $\Omega_{\Lambda} = 0.7$ cosmology throughout this paper.
\begin{figure*}
	\centering
	\includegraphics[scale=0.39]{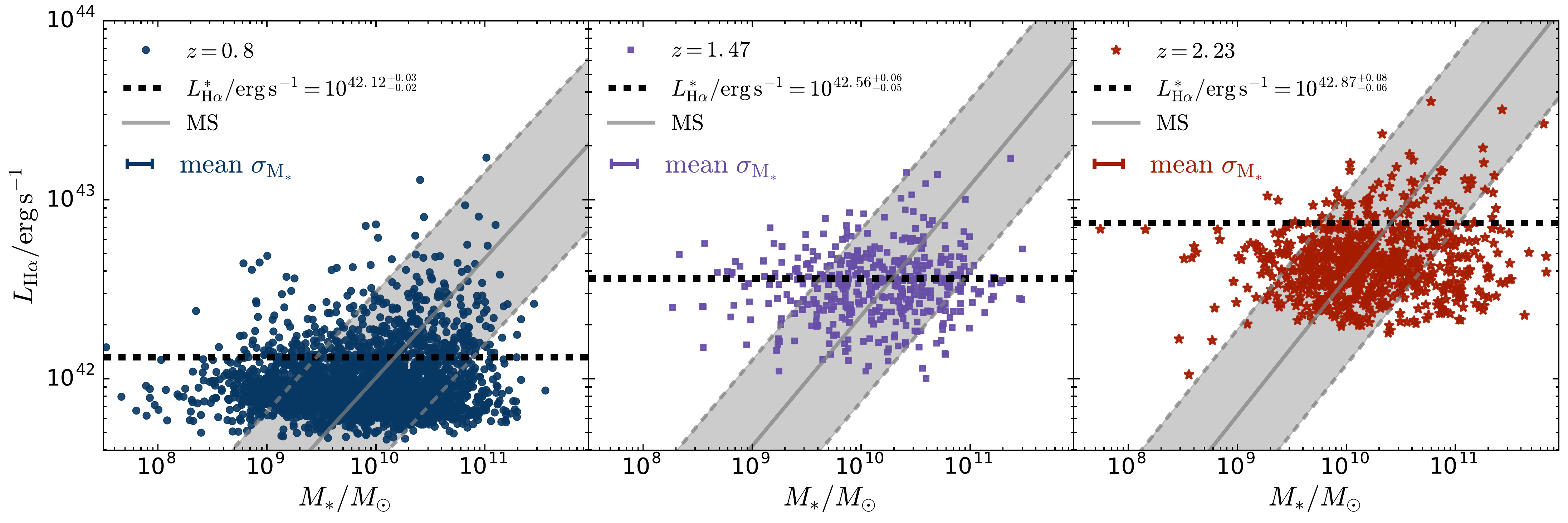}	
	\caption{Distributions of SED-estimated stellar masses and dust-corrected $\rm{H}\alpha$ luminosities for the three samples of HiZELS galaxies, at $z=0.8$, $z=1.47$ and $z=2.23$. The dashed lines show $L_{\rm{H}\alpha}^{*}$ at each redshift, derived by \protect\cite{Sobral2012} and \protect\cite{Cochrane2017}. Overplotted are indicative regions of the `main sequence' at each redshift with $2\sigma$ contours, derived by \protect\cite{Speagle2014}.}
	\label{fig:histogram_lum_mass}
\end{figure*}
\section{The HiZELS survey and sample selection}\label{sec:sample_selection}
\subsection{Samples of $\rm{H}\alpha$ emitters}\label{sec:ha_sample}
Our sources are drawn from HiZELS, selected by their emission line strength as detailed in \cite{Sobral2012} and \cite{Sobral2015}. A combination of narrow- and broad-band images are used to identify $\rm{H}\alpha$ emitters, yielding sources within narrow redshift ranges ($\Delta z\sim0.02$) centred on $z=0.81 \,\& \, 0.84$ (hereafter $z=0.8$), $z=1.47$, $z=2.23$. The galaxies used in this paper are the same as those used by C17: we impose the criterion that sources exceed $f_{50}$, the 50\% completeness flux of their survey frames. Raw $\rm{H}\alpha$ narrow-band fluxes are corrected for dust extinction by $0.4\,\rm{dex}$ ($\rm{A}_{\rm{H}\alpha}=1$). An equivalent width-dependent $[\rm{NII]}$ line contamination correction is made to account for emission from the $[\rm{NII]6548,\,6584}$ lines that also fall into the narrow-band filter (see \citealt{Sobral2012}). Star-formation rates are derived directly from dust-corrected $\rm{H}\alpha$ luminosities, $\rm{L}_{\rm{H}\alpha}$ using
\begin{equation}
\rm{SFR}_{\rm{H}\alpha}(M_{\odot}\rm{year}^{-1}) = 4.6 \times 10^{-42}\rm{L}_{\rm{H}\alpha}(\rm{ergs}\,s^{-1}),
\end{equation}
adopting the calibration of \cite{Jr1998} and scaling by a factor 1.7 \citep{Speagle2014} to convert from a \citet{Salpeter1955} IMF to a \citet{Chabrier2003} IMF.
\begin{table}
\begin{center}
\begin{tabular}{l|c|c}
Field & $\bar{z}_{\rm{H}\alpha\,\,\rm{emitters}}$ & $\#\,\rm{H}\alpha$ emitters \\
\hline
NBJ (COSMOS \& UDS) &  $0.845\:\ensuremath{\pm}\:0.011$ & $503$ \\
NBJ (SA22)  &  $0.81\:\ensuremath{\pm}\:0.011$ & $2332$  \\
NBH (COSMOS \& UDS) &  $1.47\:\ensuremath{\pm}\:0.016$ & $451$ \\
NBK (COSMOS \& UDS) &  $2.23\:\ensuremath{\pm}\:0.016$ & $727$ \\
\end{tabular}
\caption{Numbers and mean redshifts of H\ensuremath{\alpha} emitters identified by the HiZELS survey and selected for this analysis \citep{Sobral2012,Sobral2015}. Only emitters which exceed the limiting flux, $f_{50}$, of their frames are included in this work.}
\label{Table:hizels_fp}
\end{center}
\end{table}

\subsection{Deriving stellar masses from deep broad-band imaging}\label{sec:full_sed}
\begin{figure}
	\centering
	\includegraphics[scale=0.46]{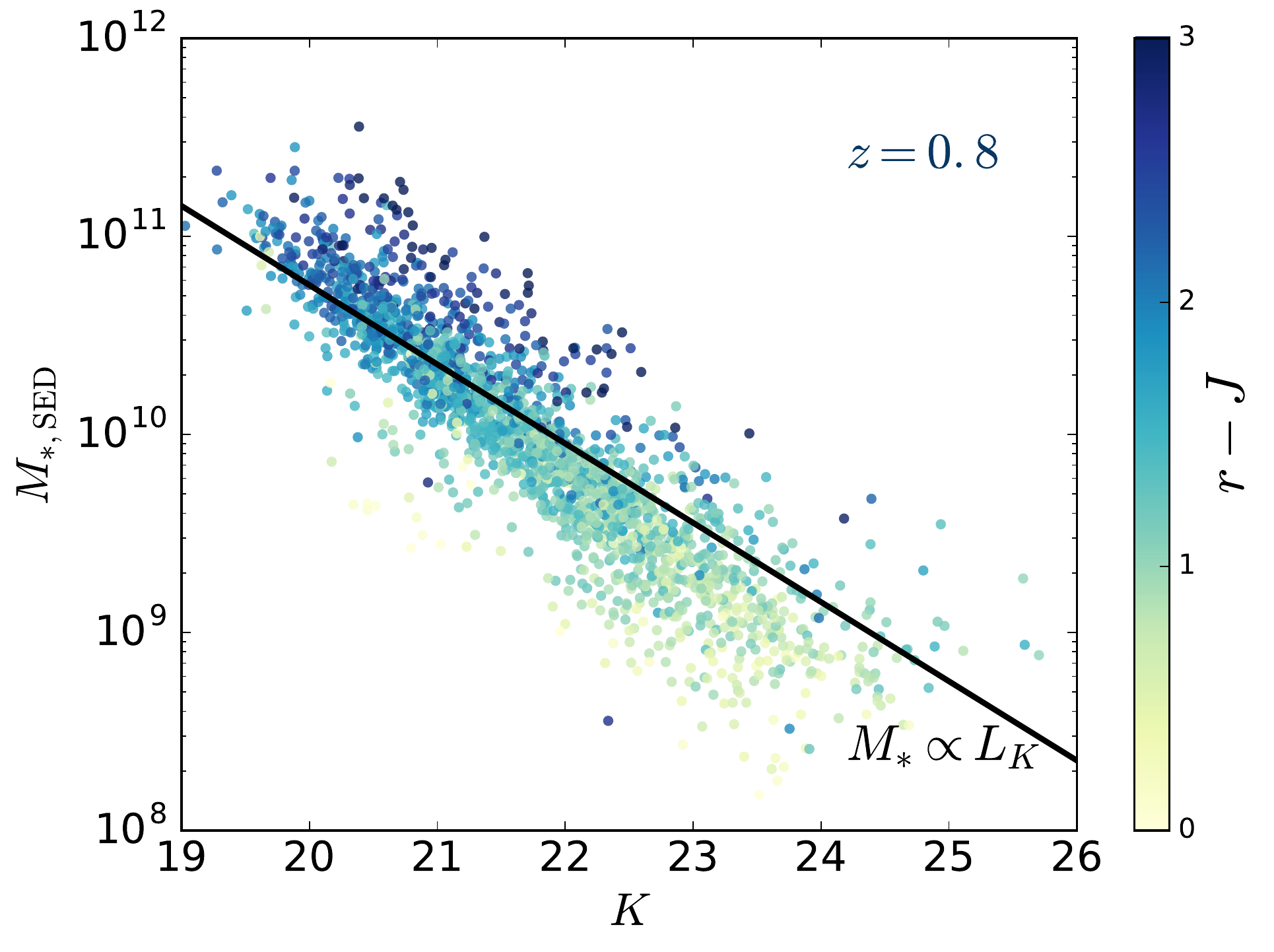}	
	\caption{SED-derived stellar mass versus observed $K$-band magnitude for SA22 galaxies, colour-coded by $r-J$ colour. The black line shows the direct proportionality between $K$-band flux (rest-frame $1.2\mu m$) and stellar mass (i.e. gradient fixed at $-0.4$). The stellar mass is clearly well correlated with $K$-band flux, but at fixed $K$-band magnitude, redder galaxies have higher SED-derived stellar masses, as would be expected. This colour dependence appears to drive the scatter in the relation and the deviation of the points from the straight line shown.}
	\label{fig:mag-k}
\end{figure}

In order to estimate stellar mass, we model each galaxy's stellar populations and dust content via spectral energy distribution (SED) fitting using a similar method to that described in \cite{Sobral2011} and \cite{Sobral2014}. The observed photometry is first shifted into the rest-frame. Model galaxy SEDs are then convolved with the detector's spectral response function to compare modelled and observed flux, and fitted via $\chi^{2}$ minimization. \\
\indent Our modelling draws upon the stellar population synthesis package of \citet{Charlot2003}, using the updated models commonly referred to as CB07. These models assume a \citet{Chabrier2003} IMF and an exponentially declining star-formation history of the form $e^{-t/\tau}$, where $\tau$ is in the range $0.1-10\rm{Gyr}$. Although this is not a realistic description of the star-formation histories of individual galaxies, which are likely to be  characterized by shorter bursts, triggered by stochastic accretion, $\tau$ is a reasonable estimate of the mean age of a galaxy (see also \citealt{Sobral2014}, who show that using single exponential star-formation models does not introduce any significant bias into the stellar mass estimates of HiZELS galaxies). We use a grid of ages from $30$Myr to the age of the Universe at each redshift, with a grid of dust extinctions from \citet{Calzetti2000} up to $\rm{E(B-V)}=0.5$, and three metallicities ($0.2-1.0 Z_{\odot}$).\\
\indent For the COSMOS field, up to 36 wide, medium and narrow bands are used, from {\it{GALEX}}'s far-UV band to {\it{Spitzer}}'s four IRAC bands. In the UDS field there are only 16 available bands, but $J$, $H$ and $K$ data from UKIRT/UKIDSS DR5 are very deep. Seven bands ($ugrizJK$) are used in SA22 \citep[see][]{Sobral2013}. All HiZELS sources are assumed to lie at the central wavelength of the redshift distribution, which is a reasonable approximation since the filter profile is extremely narrow (see Table \ref{Table:hizels_fp}). The resultant stellar masses are fairly well constrained, with typical statistical uncertainties of $0.23,\,0.24$ and $0.26\,\rm{dex}$ at $z=0.8,\,1.47$ and $2.23$, which vary a little from source to source. SED masses are plotted against $\rm{H}\alpha$ luminosities for the HiZELS samples in Figure \ref{fig:histogram_lum_mass}. At each redshift, our samples cover a very wide range in stellar mass ($10^{8}<M_{*}/M_{\odot}<10^{11}$) and also around $1\,\rm{dex}$ in $\rm{H}\alpha$ luminosity, spanning the break of the luminosity function.
\\
\indent As a test of our stellar masses, especially in SA22, where fewer bands are available, we compare our stellar mass estimates to apparent $K$-band luminosities, which broadly trace the older stellar population \citep[e.g.][]{Kauffmann1998,Longhetti2009}. Figure \ref{fig:mag-k} shows SED-derived stellar mass versus observed $K$-band magnitude for HiZELS galaxies in the SA22 field at $z=0.8$. These galaxies occupy a clear locus in this plane, close to the line expected from direct proportionality between $K$-band flux (rest-frame $1.2\mu m$) and stellar mass. At fixed $K$-band magnitude, redder galaxies (see colour coding) have higher SED masses than would be expected from a naive extrapolation of $K$-band flux, and bluer galaxies have lower derived SED masses. This is exactly as expected, since the red fraction is higher for higher luminosity sources. These galaxies are dominated by old stars and have high mass-to-light ratios. In contrast, the bluer (typically less luminous) galaxies in our HiZELS samples have younger stellar populations, and are thus particularly luminous for their mass. We conclude that our SED masses are reasonable, and fold in important colour information. Therefore, we use the SED-derived stellar masses for the remainder of this paper, with confidence. We note, nevertheless, that our results are qualitatively unchanged whether we use $K$-band-derived or SED-derived masses. \\
\section{Quantifying galaxy clustering using the two-point correlation function}\label{sec:clustering_one}
We quantify the clustering of subsamples of HiZELS galaxies using the same techniques as C17, and the interested reader should refer to that paper for more details. Here, we provide a brief overview of our methods.
\begin{figure*} 
	\centering
	\includegraphics[scale=0.5]{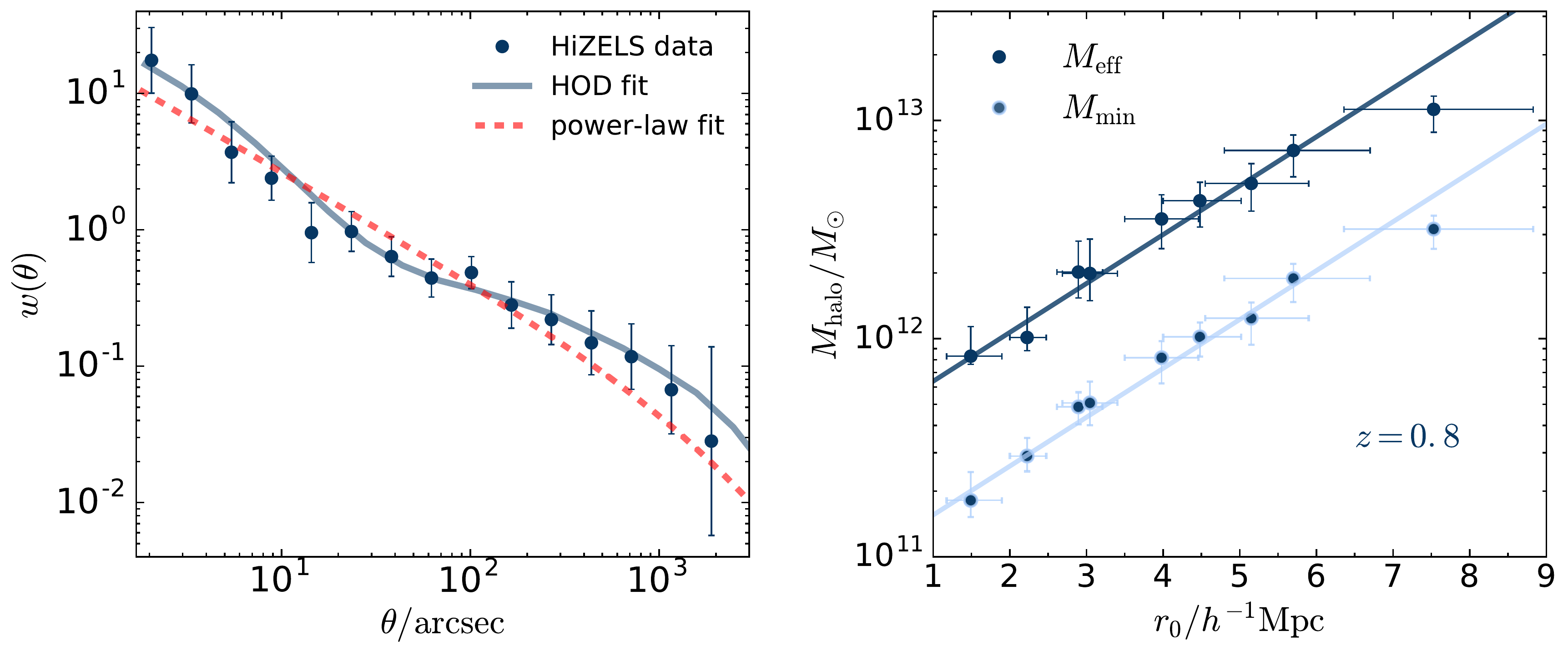}
	\caption{Left: The two-point angular correlation function constructed for the whole sample at $z=0.8$, fitted with a power-law ($r_{0}=2.58^{+0.16}_{-0.14}h^{-1}\rm{Mpc}$) and HOD model ($M_{\rm{eff}}=12.13^{+0.10}_{-0.09} M_{\odot}$). Right: $r_{0} - M_{\rm{halo}}$ calibration from \protect\cite{Cochrane2017}. Overplotted are the best-fitting relations $\log_{10}M_{\rm{eff}}/M_{\odot} =  11.7 \pm 0.7 + r_{0}/(4.5\pm0.3)$ and $\log_{10}M_{\rm{min}}/M_{\odot} = 10.9 \pm 0.7 + r_{0}/(4.5\pm0.3)$. We find excellent linear fits, so use $r_{0}$ as a proxy for halo mass in this paper.}
    \label{fig:calibrating_r0_fig}
\end{figure*}
\subsection{Angular two-point clustering statistics}\label{sec:two_point}
The angular two-point correlation function, $w(\theta)$, is defined as the excess probability of finding a pair of galaxies separated by a given angular distance, relative to that probability for a uniform (unclustered) distribution with the same areal coverage. The probability $dP(\theta)$ of finding galaxies in solid angles $d\Omega_{1}$ and $d\Omega_{2}$ is thus $dP(\theta) = N^{2}(1+w(\theta))\,d\Omega_{1}d\Omega_{2}$, where $N$ is the surface density of galaxies. $w(\theta)$ is generally calculated by comparing the distribution of sources to that of a randomly distributed population subject to the same sample selection criteria. We use random samples of galaxies as described in C17. Random galaxies have luminosities drawn from the luminosity function constructed from the same samples, not exceeding the limiting flux of their simulated detection frame, and taking into account the effects of incompleteness and flux boosting. \\
\indent Following C17, we use the minimum variance estimator proposed by \cite{Landy1993}, which was shown to be minimally susceptible to bias from small sample sizes and fields:
\begin{equation} \label{eq:w_theta}
w(\theta) = 1 + \left(\frac{N_R}{N_D}\right)^{2}\frac{DD(\theta)}{RR(\theta)} - 2\frac{N_{R}}{N_{D}}\frac{DR(\theta)}{RR(\theta)}.
\end{equation}
$N_{R}$ and $N_{D}$ are the total number of random and data galaxies in the sample, and $RR(\theta)$, $DD(\theta)$, and $DR(\theta)$ correspond to the number of random-random, data-data, and data-random pairs separated by angle $\theta$. $w(\theta)$ is normally fitted with a power-law, $w(\theta) = A\theta^{-0.8}$. \\
\indent We estimate uncertainty using the bootstrap resampling method, with the HiZELS observed frames forming our resampled volumes. Each correlation function was constructed from 1000 bootstraps, taking the error on each $w(\theta)$ bin as the diagonal element of the bootstrap covariance matrix. These uncertainties are quite conservative \citep[see][]{Norberg2009}, enhanced by variations between frames of different depths. As described in C17, we make a small correction, the integral constraint \citep{Groth1977}, to account for the underestimation of clustering strength due to the finite area surveyed. 
\subsection{Obtaining a real-space correlation length}\label{sec:correlation_length}
In order to compare the clustering strengths of populations of star-forming galaxies at different redshifts quantitatively, we convert the angular correlation function to a spatial one. This conversion is often performed using Limber's approximation \citep{Limber1953}, which assumes that spatial correlations that follow $\xi = (r/r_{0})^{\gamma}$ are projected as angular correlation functions with slopes $\beta=\gamma+1$. This simple power-law fit is not reliable for our samples of galaxies, which span fields with separations of degrees and use very narrow filters, meaning that on large scales, the angular separation directly traces the real-space separation (resulting in a slope $\beta=\gamma$ on large scales). Therefore, we perform a numerical integration of the exact equation:
\begin{equation}\label{eq:filter_integration}
w_{\rm{model}}(\theta) = \psi^{-1}\int_{0}^{+\infty}\int_{s\sqrt{2\phi}}^{2s}\frac{2f_{s}(s-\Delta)f_{s}(s+\Delta)}{R^{-\gamma-1}r_{0}^{\gamma}\Delta}dRds.{}
\end{equation}\\{}
Here, $\psi=1+\cos\theta$, $\phi=1-\cos\theta$, $\Delta=\sqrt{(R^{2}-2s^{2}\phi)/2\psi}$, and $f_{s}$ is the profile of the filter, fitted as a Gaussian profile with $\mu$ and $\sigma$ that depend on the filter being considered (see C17 for the parameters of our filters). We assume the standard value of $\gamma = -1.8$. $\chi^{2}$ fitting of observed against modelled $w(\theta)$, generated using different $r_{0}$ values, allows us to estimate $r_{0}$ and its error \citep[see][]{Sobral2010}.
\subsection{Halo Occupation Distribution fitting to obtain halo masses}\label{sec:halo_fitting}
In C17, we used Halo Occupation Distribution (HOD) modelling to derive typical dark matter halo masses for $\rm{H}\alpha$ luminosity-binned samples of HiZELS galaxies. HOD modelling involves parametrizing the number of galaxies per halo as a function of dark matter halo mass, $\langle N|M\rangle$. Given a set of HOD parameters, a halo mass function and halo bias (here both are adopted from \citealt{Tinker2010}) and a halo profile (we use NFW; \citealt{Navarro1996}) we generate a real-space correlation function. For each parameter instance, we simulate the projection of this real-space correlation function and compare the result to our observed two-point correlation functions. We use Markov chain Monte Carlo (MCMC) techniques, implemented using the {\small{EMCEE}} package \citep{Foreman-Mackey2013}, to determine the best-fitting parameters. All fitting is performed using the HMF and HALOMOD packages provided by \cite{Murray2013}. \\
\indent Satellite galaxies are parametrized to have a power-law occupancy above some halo mass, in line with most HOD models. The HOD parametrization of centrals differs from those formulated for mass-limited samples, because although all massive haloes will contain a central galaxy, this need not fall within a star-formation rate limited sample. Recent work by \cite{Gonzalez-Perez2017} supports adopting an alternative parametrization for star-forming galaxies, which includes a Gaussian peak for low-mass haloes. Thus, following \cite{Geach2012} and C17, we parametrize the number of central and satellite galaxies separately as: 
\begin{equation}
\begin{split}
\langle N_{\rm{cen}}|\rm{M}\rangle = F_{c}^{B}(1-F_{c}^{A})\rm{exp}\Bigg[-\frac{\log(M/M_{min})^{2}}{2(\sigma_{\log M})^{2}}\Bigg] \\
+ \frac{1}{2}F_{c}^{A}\Bigg[1+\rm{erf}\Bigg(\frac{\log(M/M_{min})}{\sigma_{\log M}}\Bigg)\Bigg],
\end{split}
\end{equation}
\begin{equation}
\langle N_{\rm{sat}}|\rm{M}\rangle = F_{s}\Bigg[1+\rm{erf}\Bigg(\frac{\log(M/M_{\rm{min}})}{\sigma_{\log M}}\Bigg)\Bigg]\Bigg(\frac{M}{M_{\rm{min}}}\Bigg)^{\alpha}.
\end{equation}
The key parameters are:
\begin{itemize}[leftmargin=0.3cm]
  \item[--] $M_{\rm{min}}$: the minimum halo mass that hosts a galaxy. Note that our definition differs subtly to that used in work characterizing mass-limited samples, such as \cite{Mccracken2015} and \cite{Hatfield2015}, since in this work $M_{\rm{min}}$ applies to both central and satellite galaxies.
  \item[--] $\sigma_{\log M}$: characterises the width of the transition to  $\langle N_{\rm{sat}}|M\rangle = F_{s} \Big(\frac{M}{M_{\rm{min}}}\Big)^{\alpha}$ around $M_{\rm{min}}$.
  \item[--] $\alpha$: the slope of the power-law for $\langle N_{\rm{sat}}|M\rangle$ in haloes with $M>M_{\rm{min}}$. In line with the literature, we fix $\alpha=1$. Tests allowing $\alpha$ to vary confirm that this is an appropriate choice.
  \item[--] $F_{c}^{A,B}$: normalization factors, in range [0,1].
  \item[--] $F_{s}$: the mean number of satellite galaxies per halo, at $M = M_{\rm{min}}$
  \end{itemize}
The total number of galaxies is given by:
\begin{equation}
\langle N|M\rangle = \langle N_{\rm{cen}}|M\rangle + \langle N_{\rm{sat}}|M\rangle.
\end{equation}
When fitting the models to data, we use the observed number density of galaxies as an additional constraint. For a given $\langle N|M\rangle$ output from the halo model, the predicted number density of galaxies is:
\begin{equation}
n_{g} =\int dM n(M) \langle N|M\rangle,
\end{equation}
where $n(M)$ is the halo mass function, for which we use the determination of \cite{Tinker2010}. The observed number density of galaxies used here is the integral of the luminosity function between the same limits used to select the real and random galaxy sample. \\
\indent For each set of HOD parameters, we may derive a number of parameters of interest for galaxy evolution. In this paper, we use the effective halo mass, the typical mass of galaxy host halo. This is given by:
\begin{equation}
M_{\rm{eff}} = \frac{1}{n_{g}}\int dM M n(M) \langle N|M\rangle.
\end{equation}
\begin{figure*} 
	\centering  
	\includegraphics[scale=0.45]{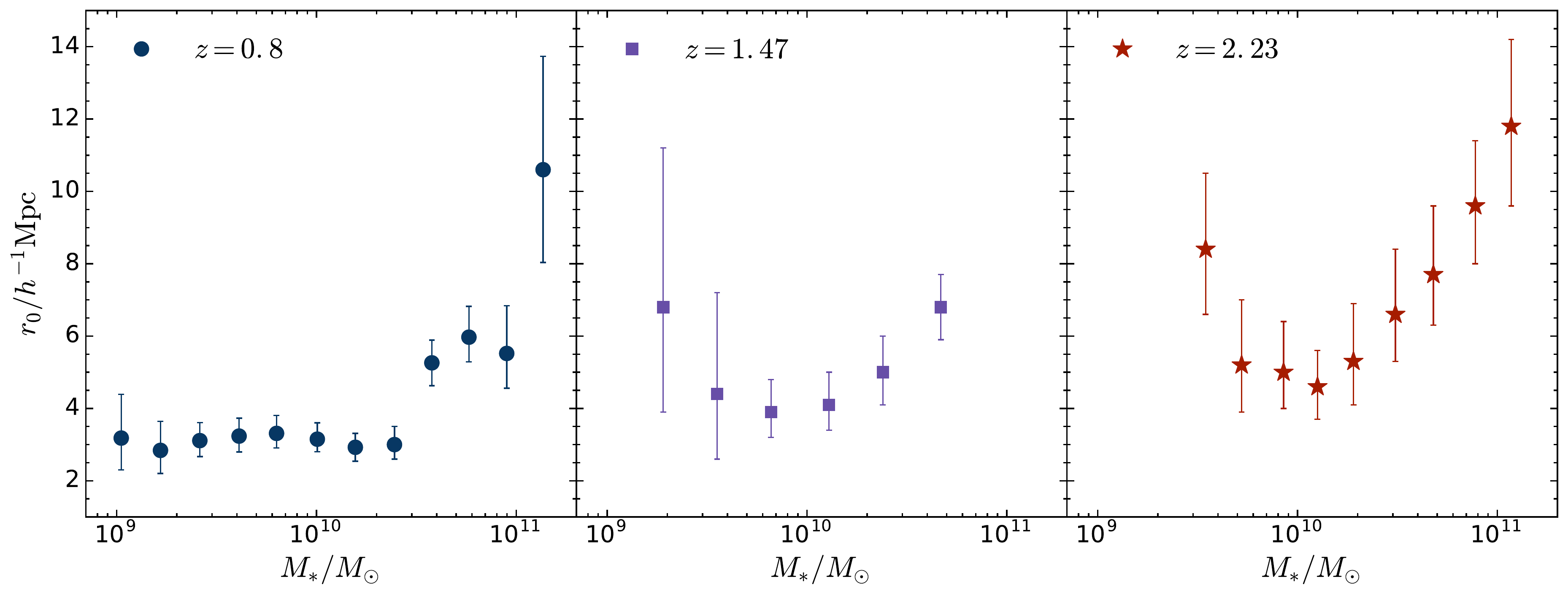}
	\includegraphics[scale=0.45]{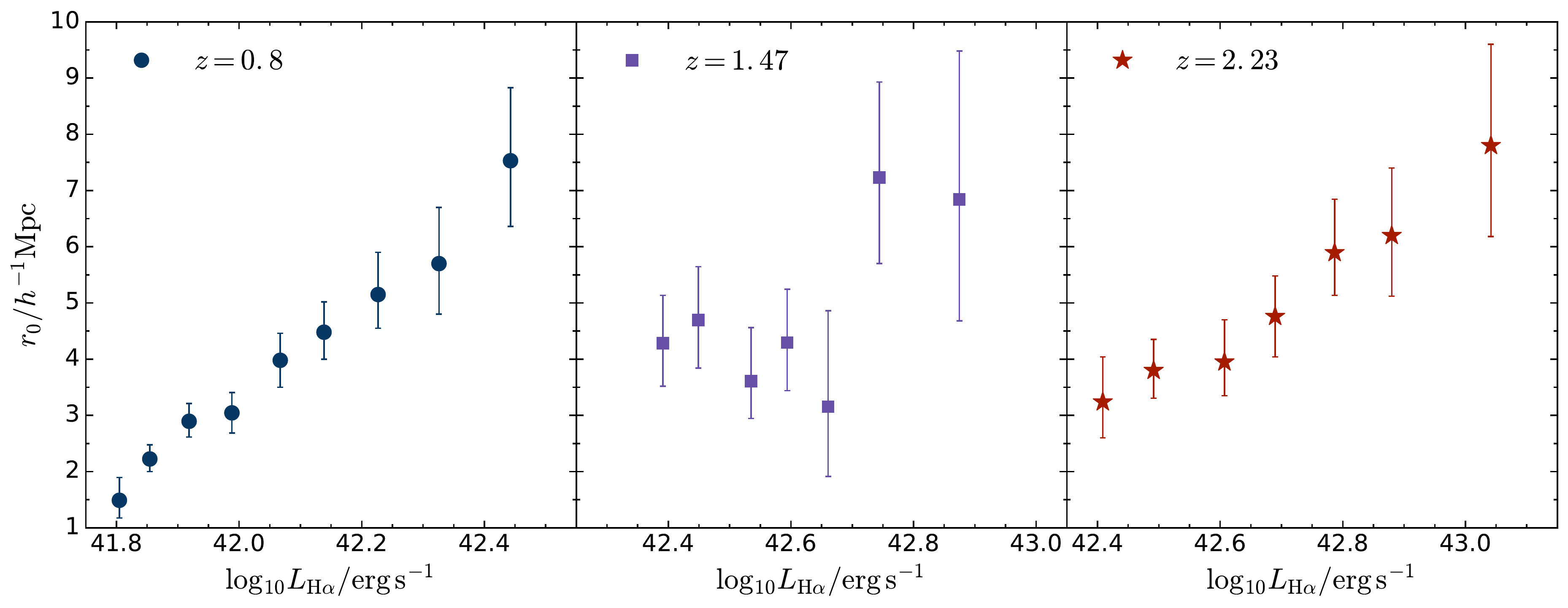}
	\caption{Top: clustering strength, $r_{0}$, as a function of stellar mass. At all three redshifts, the clustering strength is broadly flat at low stellar masses, with evidence for an increase for the most massive galaxies (above $\sim2-3\times10^{10}M_{\odot}$). Bottom: $r_{0}$ versus $L_{\rm{H}\alpha}$ from C17, replotted for comparison. Here, a strong monotonic trend is seen between $r_{0}$ and $L_{\rm{H}\alpha}$ at $z=0.8$ and $z=2.2$; as shown in C17, the $z=1.47$ data are consistent with the same trend (albeit noisier due to the smaller sample).}
    \label{fig:r0_mass_or_lum}
\end{figure*}	
\subsection{Calibrating $r_{0}$ to $M_{\rm{halo}}$ using HOD models}\label{sec:hod_cal}
For samples of galaxies with large satellite fractions, there will be a substantial one-halo term in the correlation function at small separations. In such cases, HOD modelling offers a better fit than a simple power-law. In C17, we found that HiZELS samples at $z=0.8$, $z=1.47$ and $z=2.23$ have low satellite fractions ($\sim5\%$), and HOD fitting offers only marginal gains in goodness of fit at small scales (see Figure \ref{fig:calibrating_r0_fig}, left-hand panel). Instead, the main benefit of HOD fitting is to allow the conversion of clustering strengths into typical halo masses. Comparing measured $r_{0}$ to derived halo masses (Figure \ref{fig:calibrating_r0_fig}, right-hand panel), we find that these are tightly correlated, and can be reasonably approximated as simple linear fits. At $z=0.8$, these are given by:
\begin{equation}
\log_{10}M_{\rm{eff}}/M_{\odot} = 11.7 \pm 0.7 + r_{0}/(4.5\pm0.3)
\end{equation}
\begin{equation}
\log_{10}M_{\rm{min}}/M_{\odot} = 10.9 \pm 0.7 + r_{0}/(4.5\pm0.3).
\end{equation}
Therefore, in some parts of this paper (Section \ref{sec:results1} - \ref{sec:compare_mass}), we simply derive and quote $r_{0}$ values, as these are sufficient to indicate trends of clustering with stellar mass or star-formation rate. When we require robust halo masses, as in Sections \ref{sec:moster_predictions} and \ref{sec:eagle_comparison}, we perform the full HOD fitting. 
\section{Clustering of HiZELS galaxies as a function of stellar mass and $\rm{SFR}$}\label{sec:results}
\subsection{Clustering as a function of $\rm{H}\alpha$ luminosity}\label{sec:results1}
In C17, we studied the clustering of HiZELS galaxies as a function of their $\rm{H}\alpha$ luminosity. We found strong relationships between $L_{\rm{H}\alpha}$ and $r_{0}$. The clustering strength increases monotonically with $\rm{H}\alpha$ luminosity at all redshifts, indicating that the most highly star-forming galaxies thrive in higher dark matter overdensities (see Figure \ref{fig:r0_mass_or_lum}). We speculated that this is where a plentiful gas supply fuels high star-formation rates. \\
\indent HOD fitting revealed that typical $\rm{H}\alpha$-emitting galaxies are star-forming centrals, residing in host haloes with minimum mass increasing with $\rm{H}\alpha$ luminosity from $\sim10^{11.2}M_{\odot}$ to $\sim10^{12.6}M_{\odot}$ and corresponding effective halo masses $\sim10^{11.6}M_{\odot}-10^{13}M_{\odot}$. At all three redshifts, $L_{\rm{H}\alpha}^{*}$ galaxies typically reside in haloes of effective mass $\sim10^{12}M_{\odot}$. This coincides with the halo mass predicted by theory to be maximally efficient at converting baryons into stars. Samples selected within the same $L_{\rm{H}\alpha}/L_{\rm{H}\alpha}^{*}$ range inhabit similar populations of dark matter haloes. The relationship between scaled galaxy luminosity $L_{\rm{H}\alpha}/L_{\rm{H}\alpha}^{*}$ and dark matter halo mass is largely independent of redshift.
\begin{table}
\begin{center}
\begin{tabular}{l|c|c|c|c|c}
$\log_{10}(M_{*}/M_{\odot})$ & Mean $\log_{10}(M_{*}/M_{\odot})$ & $\rm{r}_{0}/h^{-1}\rm{Mpc}$ \\
\hline
\multicolumn{2}{l}{$z=0.8$, $41.72<\log_{10}(L_{\rm{H}\alpha}/\rm{erg\,s^{-1}})<42.42$} & & & \\
\hline
$8.8-9.2$ & $9.02$ & $3.2^{+1.2}_{-0.9}$ \\
$9.0-9.4$ & $9.22$ & $2.8^{+0.8}_{-0.6}$ \\
$9.2-9.6$ & $9.42$ & $3.1^{+0.5}_{-0.4}$ \\
$9.4-9.8$ & $9.61$ & $3.2^{+0.5}_{-0.4}$ \\
$9.6-10.0$ & $9.80$ & $3.3^{+0.5}_{-0.4}$ \\
$9.8-10.2$ & $10.00$ & $3.2^{+0.5}_{-0.4}$ \\
$10.0-10.4$ & $10.19$ & $2.9^{+0.4}_{-0.4}$ \\
$10.2-10.6$ & $10.39$ & $3.0^{+0.5}_{-0.4}$ \\
$10.4-10.8$ & $10.58$ & $5.3^{+0.6}_{-0.6}$ \\
$10.6-11.0$ & $10.76$ & $6.0^{+0.9}_{-0.7}$ \\
$10.8-11.2$ & $10.95$ & $5.5^{+1.3}_{-1.0}$ \\
$11.0-11.4$ & $11.13$ & $10.6^{+3.1}_{-2.6}$ \\
\hline
\multicolumn{2}{l}{$z=1.47$, $42.16<\log_{10}(L_{\rm{H}\alpha}/\rm{erg\,s^{-1}})<42.86$} & & & \\
\hline
$8.9-9.5$ & $9.28$ & $6.8^{+4.4}_{-2.9}$ \\
$9.2-9.8$ & $9.55$ & $4.4^{+2.8}_{-1.8}$ \\
$9.5-10.1$ & $9.82$ & $3.9^{+0.9}_{-0.7}$ \\
$9.8-10.4$ & $10.11$ & $4.1^{+0.9}_{-0.7}$ \\
$10.1-10.7$ & $10.38$ & $5.0^{+1.0}_{-0.9}$ \\
$10.4-11.0$ & $10.67$ & $6.8^{+1.1}_{-0.9}$ \\
\hline
\multicolumn{2}{l}{$z=2.23$, $42.47<\log_{10}(L_{\rm{H}\alpha}/\rm{erg\,s^{-1}})<43.17$} & & & \\
\hline
$9.3-9.7$ & $9.54$ & $8.4^{+2.1}_{-1.8}$ \\
$9.5-9.9$ & $9.72$ & $5.2^{+1.8}_{-1.3}$ \\
$9.7-10.1$ & $9.93$ & $5.0^{+1.4}_{-1.0}$ \\
$9.9-10.3$ & $10.10$ & $4.6^{+1.0}_{-0.9}$ \\
$10.1-10.5$ & $10.28$ & $5.3^{+1.6}_{-1.2}$ \\
$10.3-10.7$ & $10.49$ & $6.6^{+1.8}_{-1.3}$ \\
$10.5-10.9$ & $10.68$ & $7.7^{+1.9}_{-1.4}$ \\
$10.7-11.1$ & $10.89$ & $9.6^{+1.8}_{-1.6}$ \\
$10.9-11.3$ & $11.07$ & $11.8^{+2.4}_{-2.2}$ \\
\end{tabular}
\caption{Clustering strength, $r_{0}$, for stellar mass-binned samples of HiZELS galaxies at $z=0.8,\,1.47$, and $2.23$.}
\label{table:r0_table}
\end{center}
\end{table}
\begin{figure}
	\centering
	\includegraphics[scale=0.48]{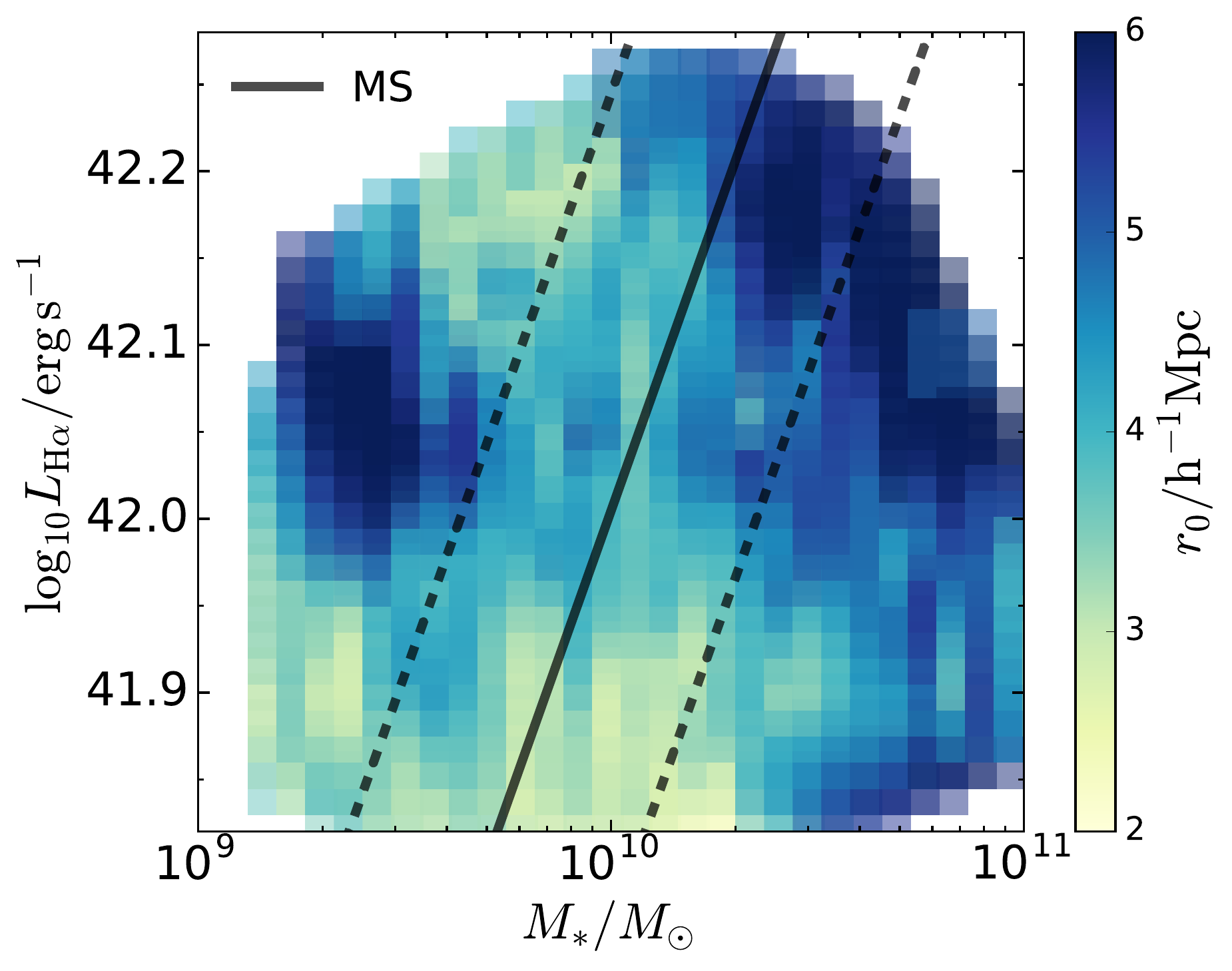}
	\caption{$r_{0}$ in the stellar mass - $L_{\rm{H}\alpha}$ plane at $z=0.8$, constructed using $\sim500$ overlapping (non-independent) subsamples and plotted using a smoothed linear interpolation. We overplot the main sequence derived by \protect\cite{Speagle2014} at this redshift as a solid line, with the dashed lines showing the standard deviation. Clustering strength increases broadly monotonically with $L_{\rm{H}\alpha}$ at all stellar masses. At high stellar masses $M_{*}\gtrsim2\times10^{10}M_{\odot}$, $r_{0}$ increases with stellar mass. We also find large $r_{0}$ values for highly star-forming low stellar mass galaxies that are located well above the main sequence.}
	\label{fig:ali_2dplot}
\end{figure}
\subsection{Clustering as a function of stellar mass}\label{sec:clustering_vs_sm}\label{sec:results2}
C17 briefly looked at $K$-band observed luminosities. We found that the trends in clustering strength with $L_{\rm{H}\alpha}$ do not differ between two large $K$-band bins, concluding that they are unlikely to be driven by stellar mass. Here, we extend that study to provide a more definitive answer to the role stellar mass plays. \\
\indent Initially we bin our sample of $z\sim0.8$ HiZELS galaxies by stellar mass, construct correlation functions and fit these as described in Section \ref{sec:two_point}, obtaining a clustering strength $r_{0}$ for each subsample. We use the broad bins in $\rm{H}\alpha$ luminosity as defined by C17 ($-0.4<\log_{10}(L_{\rm{H}\alpha}/L_{\rm{H}\alpha}^{*})<0.3$) for consistency, but find no significant differences when we re-run the analysis with no luminosity cuts except for the HiZELS selection. We find that the clustering strength is broadly constant with stellar mass at low galaxy masses. This is particularly clear at $z=0.8$, where our samples are largest and probe lowest in stellar mass, but all three redshifts are consistent with this result. The clustering strength only increases when we reach stellar mass bins that contain a significant number of galaxies below the main sequence: at all three HiZELS redshifts, clustering strength increases significantly above a mass $2-3\times10^{10}M_{\odot}$ and the most massive galaxies are very strongly clustered (see Figure \ref{fig:r0_mass_or_lum} and Table \ref{table:r0_table}). For our $\rm{H}\alpha$-selected samples, the $M_{*}-r_{0}$  relationship appears substantially weaker than the $L_{\rm{H}\alpha}-r_{0}$ relation obtained by C17, and shown in Figure \ref{fig:r0_mass_or_lum} for comparison, which continues to decrease at low $\rm{H}\alpha$ luminosities.\\ 
\indent Whilst the gradient of the stellar mass - halo mass relation of mass-selected galaxies does decrease below $M_{*}\sim10^{10}M_{\odot}$ (see Section \ref{sec:moster_predictions}; \citealt{Moster2010,Moster2013,Behroozi2013} and many others), the flattening we observe for these $\rm{H}\alpha$-selected galaxies is very pronounced. This indicates that low-mass HiZELS galaxies reside in more massive dark matter haloes than would be expected for star-forming central galaxies of these stellar masses. Although this might be surprising, given that C17 found low satellite fractions for these samples, it is important to remember that, at these masses, HiZELS $\rm{H}\alpha$-selected galaxies lie well above the `main sequence'. We explore the joint dependence of clustering on both stellar mass and $L_{\rm{H}\alpha}$ in the following subsection.
\subsection{Splitting by both stellar mass and $\rm{H}\alpha$ luminosity}\label{sec:mass_AND_lum}
\begin{figure*}
	\centering
	\includegraphics[scale=0.48]{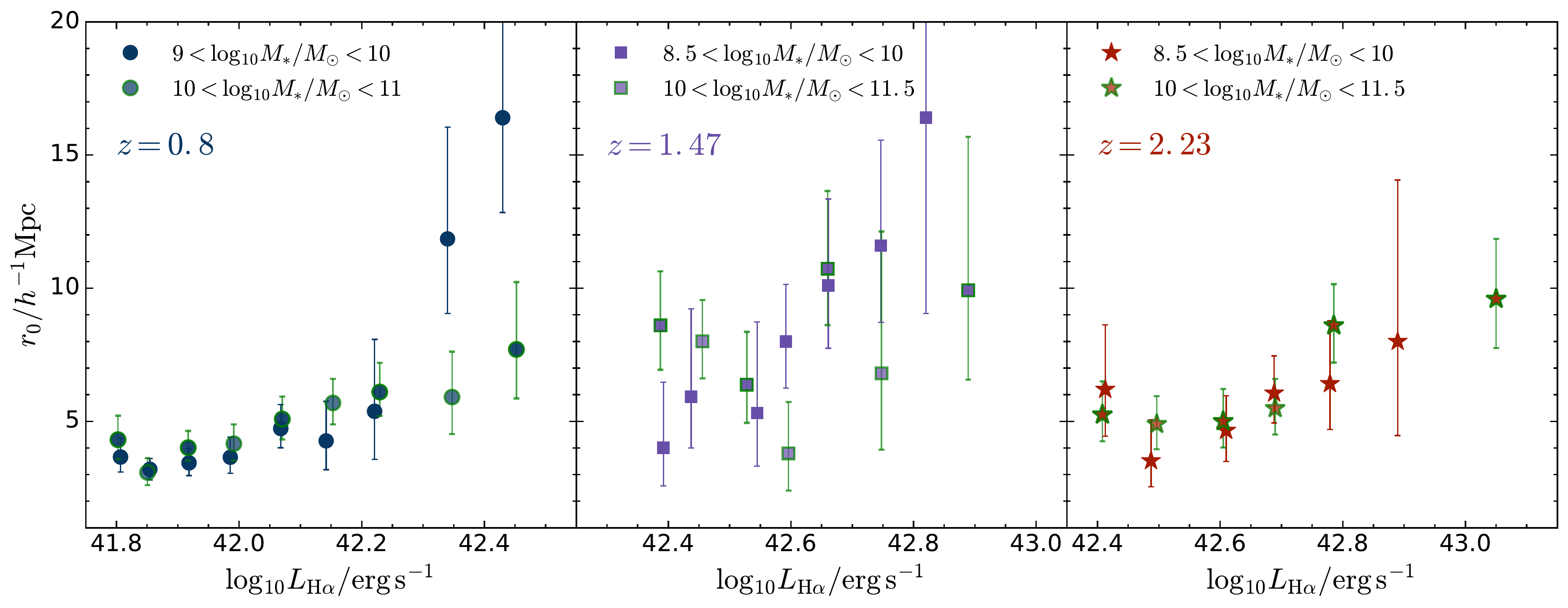}
	\caption{Clustering strength as a function of $L_{\rm{H}\alpha}$ for HiZELS galaxies split into two stellar mass bins at each redshift. The calculated $r_{0}$ values of the two mass-binned samples are consistent at fixed mass, with the possible exception of the very highest luminosities at $z=0.8$. This implies that the $\rm{H}\alpha$ luminosity is the physical property most strongly correlated with clustering strength for our HiZELS galaxies.}
	\label{fig:lum_mass_r0_comp_fig}
\end{figure*}
\begin{figure*}
	\centering
	\includegraphics[scale=0.48]{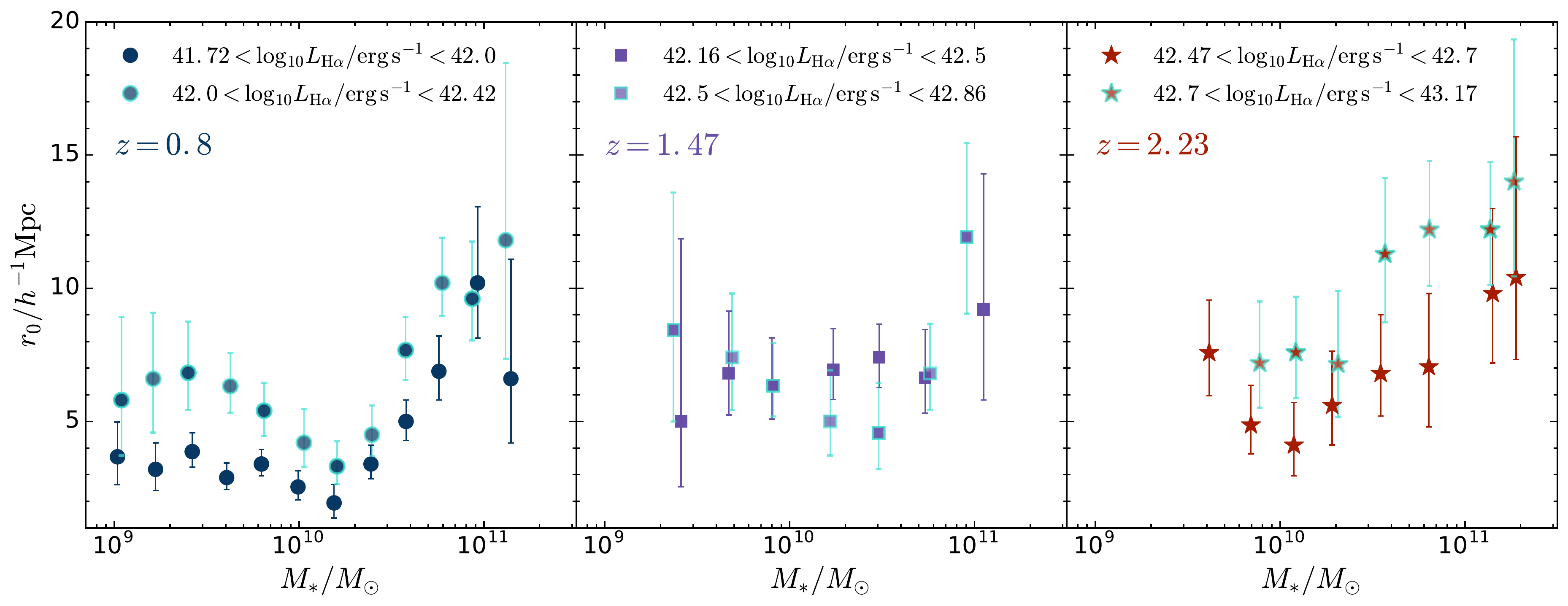}
	\caption{Clustering strength as a function of stellar mass for HiZELS galaxies split into two $\rm{H}\alpha$ luminosity bins at each redshift. Both high- and low-luminosity massive galaxies are more strongly clustered than their less massive counterparts. Higher $\rm{H}\alpha$ luminosity galaxies tend to be more strongly clustered than less luminous galaxies at fixed mass. This is clear for the two largest samples, at $z=0.8$ and $z=2.23$. The offset in $r_{0}$ between the two luminosity bins is particularly large at low stellar masses, suggesting that low-mass galaxies with high luminosities have environmentally triggered star formation.}
	\label{fig:mass_lum_r0_fig2}
\end{figure*}
\begin{figure*} 
	\centering
	\includegraphics[scale=0.48]{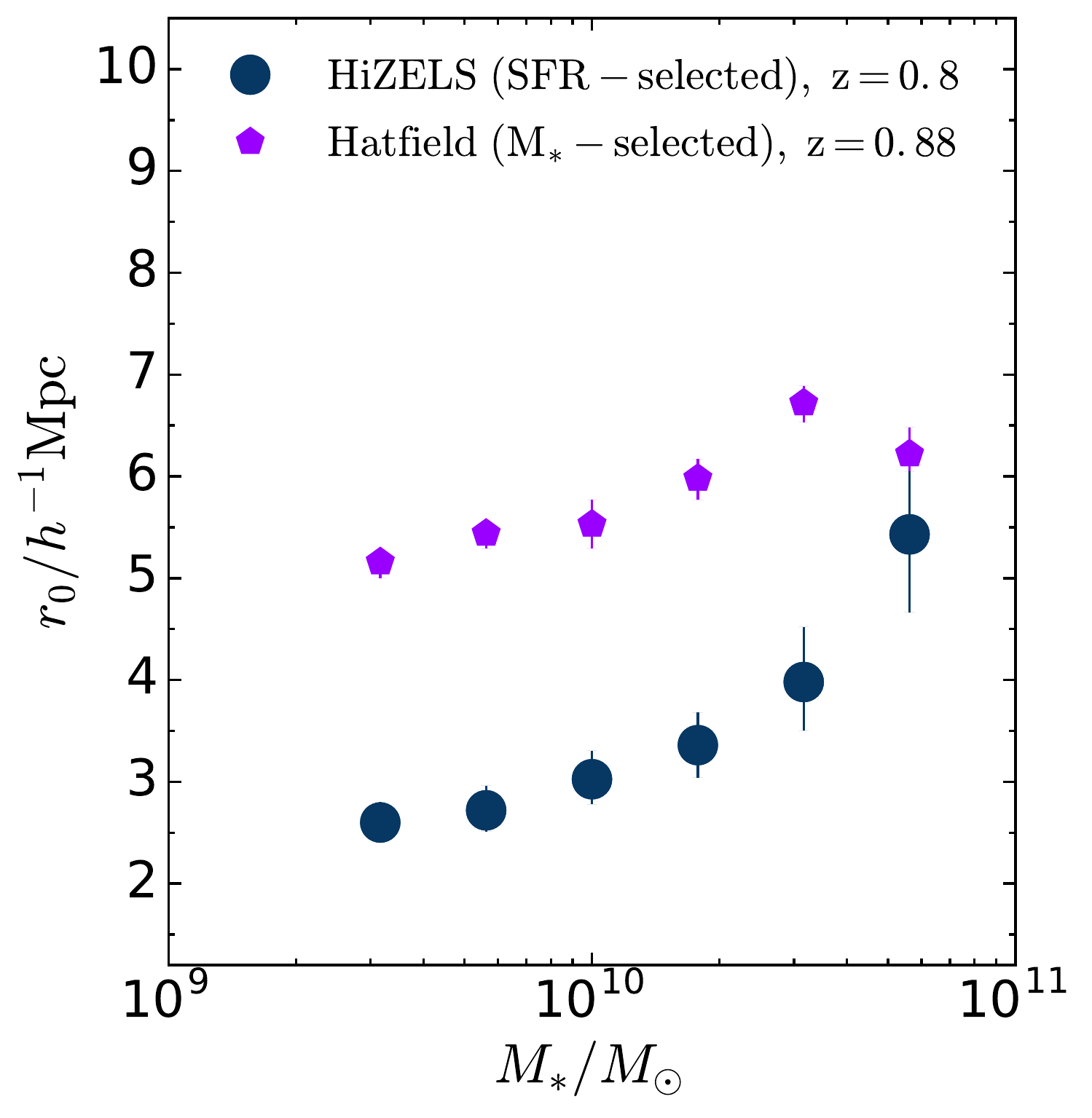}
	\includegraphics[scale=0.47]{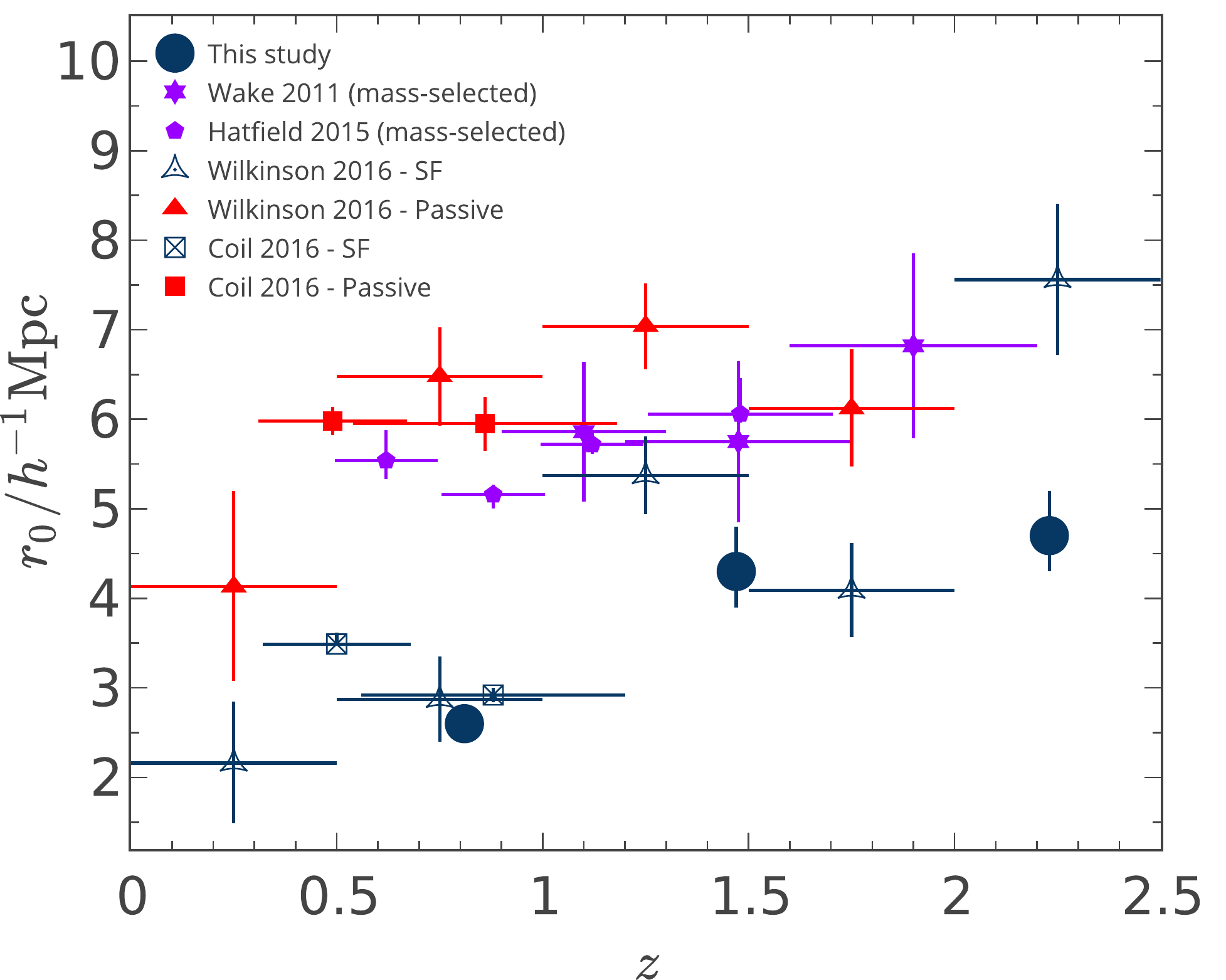}
	\caption{Left: $r_{0}$ as a function of stellar mass lower limit, for HiZELS $\rm{H}\alpha$-selected galaxies and mass-selected galaxies from \protect\cite{Hatfield2015}. At fixed stellar mass limit, the star-forming galaxies display significantly lower $r_{0}$ values, with the difference only decreasing at the highest stellar mass limits. Right: Comparison of whole-sample $r_{0}$ values at different redshifts. There are clear differences in derived $r_{0}$ due to sample selection. In general, samples of passive galaxies (red points) and mass-selected samples (purple points) tend to be more highly clustered than samples of star-forming galaxies at the same redshift (blue points).}
    \label{fig:lit_comparison}
\end{figure*}
Within the star-forming population, higher mass galaxies tend to have higher star-formation rates (and therefore higher $\rm{H}\alpha$ luminosities), so trends in mass can manifest as apparent trends in star-formation rate, and vice-versa. Here, $r_{0}$ increases significantly at both high $L_{\rm{H}\alpha}$ and high stellar masses, and it is hard to tell the extent to which mass and luminosity are each independently correlated with halo mass. Our large samples of HiZELS galaxies allow us to break this degeneracy, and study trends in stellar mass and $L_{\rm{H}\alpha}$ luminosity independently.\\
\indent At $z=0.8$, where our sample is largest, we split the stellar mass - $L_{\rm{H}\alpha}$ plane into $\sim500$ overlapping subsamples, constructing and fitting two-point correlation functions for each. In Figure \ref{fig:ali_2dplot}, we present a 2D plot of stellar mass versus $L_{\rm{H}\alpha}$. Each region is colour-coded by its $r_{0}$ value, obtained via a smoothed grid using $x$ and $y$ values of each subsample's mean stellar mass and star-formation rate, respectively. Note that these $r_{0}$ measurements are not independent, due to the overlapping samples. With around 100 galaxies per bin, there are approximately 30 independent subsamples. We find that clustering strength increases broadly monotonically with $L_{\rm{H}\alpha}$ at all stellar masses. At high stellar masses $M_{*} \ge 10^{10}M_{\odot}$, $r_{0}$ also increases with stellar mass, as has been found by many mass-selected clustering studies. At low stellar masses, the stellar mass-$r_{0}$ relationship breaks down, as had been seen in Figure \ref{fig:r0_mass_or_lum}. There is little change in $r_{0}$ with stellar mass at fixed $L_{\rm{H}\alpha}$ (if anything, $r_{0}$ increases slightly as we probe to lower stellar mass at higher $L_{\rm{H}\alpha}$, where we are probing star-formation rates well above the main sequence). \\
\indent Next, we show projections of this plot for the $z=0.8$ data, and for the smaller samples at $z=1.47$ and $z=2.23$. We divide our galaxies at each redshift slice into two stellar mass bins, and bin further by $L_{\rm{H}\alpha}$. We construct two-point correlation functions and obtain correlation strengths for these subsamples. The results are shown in Figure \ref{fig:lum_mass_r0_comp_fig}. We find that the increase in clustering strength with $\rm{H}\alpha$ luminosity holds for both stellar mass bins. The trends of the two stellar mass bins are almost indistinguishable. Only the most extremely luminous galaxies at $z=0.8$ ($L_{\rm{H}\alpha}>10^{42.2}$) show any departure from this, and, as found by \cite{Sobral2016}, HiZELS samples at these luminosities suffer from significant AGN contamination.\\
\indent We also divide our galaxies at each redshift slice into two $L_{\rm{H}\alpha}$ bins, and bin further by stellar mass. The results are shown in Figure \ref{fig:mass_lum_r0_fig2}. Given the size of the sample, our results are clearest at $z=0.8$. Here, we find that at all stellar masses, the higher luminosity galaxies are more strongly clustered than low luminosity galaxies at the same stellar mass, but this difference is most significant at low stellar masses. The data at $z=0.8$ (top panel of Figure \ref{fig:r0_mass_or_lum}) clearly shows that below stellar masses of $M_{*}\sim 10^{10}M_{\odot}$, HiZELS galaxies have a fairly flat $r_{0}$-$\rm{M}_{*}$ relation. At these stellar masses, the higher luminosity subsample displays much stronger clustering than the lower luminosity subsample, with $r_{0}\sim6-7h^{-1}\rm{Mpc}$ ($M_{\rm{eff}}\sim10^{13}M_{\odot}$), compared to $r_{0}\sim3-4h^{-1}\rm{Mpc}$ ($M_{\rm{eff}}\sim10^{12.4}M_{\odot}$). There is even a slight increase in clustering strength towards low masses for the higher luminosity subsample. We find similar trends for our second largest sample, at $z=2.23$. \\
\indent Together, our results present clear evidence for a dependence of star-formation activity of low-mass galaxies on environment. For these galaxies, $\rm{H}\alpha$ luminosity is a better predictor of clustering strength than stellar mass. As mentioned in the Introduction, the key difference between this work and many studies of galaxy clustering that use mass-selected samples is the clean, $L_{\rm{H}\alpha}$-selected sample of star-forming galaxies yielded by our survey. In order to satisfy the HiZELS $\rm{H}\alpha$ flux limit, low stellar mass galaxies must lie significantly above the main sequence. One physical interpretation of this result is that these galaxies are highly star-forming centrals, which will soon form more stellar mass to put them on the main stellar mass - halo mass relation. Alternatively, we could be observing an increasing contribution of starbursting satellite galaxies (or galaxies that are infalling on to a massive halo and will soon become satellites) at low stellar masses.
\subsection{Comparison of star-forming galaxies to mass-selected samples}\label{sec:compare_mass}
Here, we compare the clustering of our $\rm{H}\alpha$-selected samples to mass-limited samples. \cite{Hatfield2015} measure the clustering of mass-limited galaxy samples from the VIDEO survey at a very similar redshift to our $z=0.8$ sample, at $0.75<z<1.00$ with median redshift $z=0.88$.\footnote{Note that in \cite{Hatfield2015}, $r_{0}$ is not derived from a power-law fit as in this work. Instead, $r_{0}$ is defined as the radius at which the best-fitting spatial correlation function equals unity.} Their selection is based on an apparent AB magnitude limit $K_{S}<23.5$. Our observations probe slightly deeper, reaching down to $K\sim25$, but the majority of our sources also satisfy $K<23.5$. The important difference between our samples is the $\rm{H}\alpha$ flux limit of our sample. Whereas we are probing mainly the star-forming population, a substantial proportion of the \cite{Hatfield2015} sample will comprise less highly star-forming and passive galaxies. We characterize the clustering of HiZELS emitters down to the same stellar mass limits as \cite{Hatfield2015}, using no luminosity cuts other than the source selection criteria described in Section \ref{sec:ha_sample}. The results, shown in the left-hand panel of Figure \ref{fig:lit_comparison}, are strikingly different. At identical stellar mass limits, HiZELS $r_{0}$ values are approximately half of the VIDEO mass-selected sample $r_{0}$ values, with this difference only decreasing at the highest stellar masses. This shows that, at fixed stellar mass, star-forming galaxies are markedly less strongly clustered than the galaxy population as a whole. Note that for the lowest two stellar mass bins of \cite{Hatfield2015}, the $K_{S}<23.5$ selection may mean that only the reddest (and most passive, thus often most clustered) galaxies are included in the analysis, possibly biasing the points upwards relative to a fully mass-selected sample. \\
\indent We now compare the clustering of our large samples of star-forming galaxies at the three HiZELS redshifts, $z=0.8$, $z=1.47$, $z=2.23$, to other clustering measurements in the literature, to see whether these stark differences between differently selected samples persist at other redshifts. The right-hand panel of Figure \ref{fig:lit_comparison} shows the results. We find that samples of passive galaxies and mass-selected samples tend to be more highly clustered than samples of star-forming galaxies at the same redshift, to at least $z\sim2$.\\
\indent Those results form a parallel story to that already presented here. While we have studied the clustering of star-forming galaxies and shown that more highly star-forming galaxies are more strongly clustered than their less star-forming counterparts at fixed stellar mass, we show here that passive galaxies are more strongly clustered than star-forming galaxies at fixed mass. How do these two apparently contradictory results fit together? \cite{Sobral2011} show that, at fixed stellar mass for $M_{*}<10^{10.6}M_{\odot}$, the mean star-formation rate of HiZELS galaxies increases strongly with environmental overdensity ($\Sigma_{c}$) across almost the full range of overdensities probed ($2 < \Sigma_{c} < 30$), which included field galaxies and small groups. This is consistent with the main part of our study: the clustering strength of the most highly star-forming galaxies is largest. \cite{Janowiecki2017} study the atomic hydrogen gas fraction of field and small group galaxies, finding that low-mass ($M_{*} \le 10^{10.2}M_{\odot}$) galaxies in the centres of groups have gas fractions $\sim 0.3\,\rm{dex}$ higher than those in the field at fixed stellar mass. They conclude that the higher star-formation activity of these galaxies is driven by their higher gas availability. \cite{Sobral2011} also use the underlying photometric sample to estimate the star-forming fraction for HiZELS galaxies as function of overdensity. Here, the trends are different. The star-forming fraction increases slowly in the range $2 < \Sigma_{c} < 10$, but displays a sharp fall above these densities, falling to below $15\%$ in the richest clusters. This is entirely consistent with our results: the mass-selected samples of \cite{Hatfield2015} display higher clustering strengths because they are dominated by passive galaxies in richer environments, which are not detected by the HiZELS survey due to its $\rm{H}\alpha$ flux selection. This interpretation, driven by the exclusion of environmentally quenched satellites from our HiZELS samples, is in line with both the low satellite fractions found in C17, and the low $M_{\rm{eff}}$ values for HiZELS galaxies in general.
\begin{figure*}
	\centering
	\includegraphics[scale=0.55]{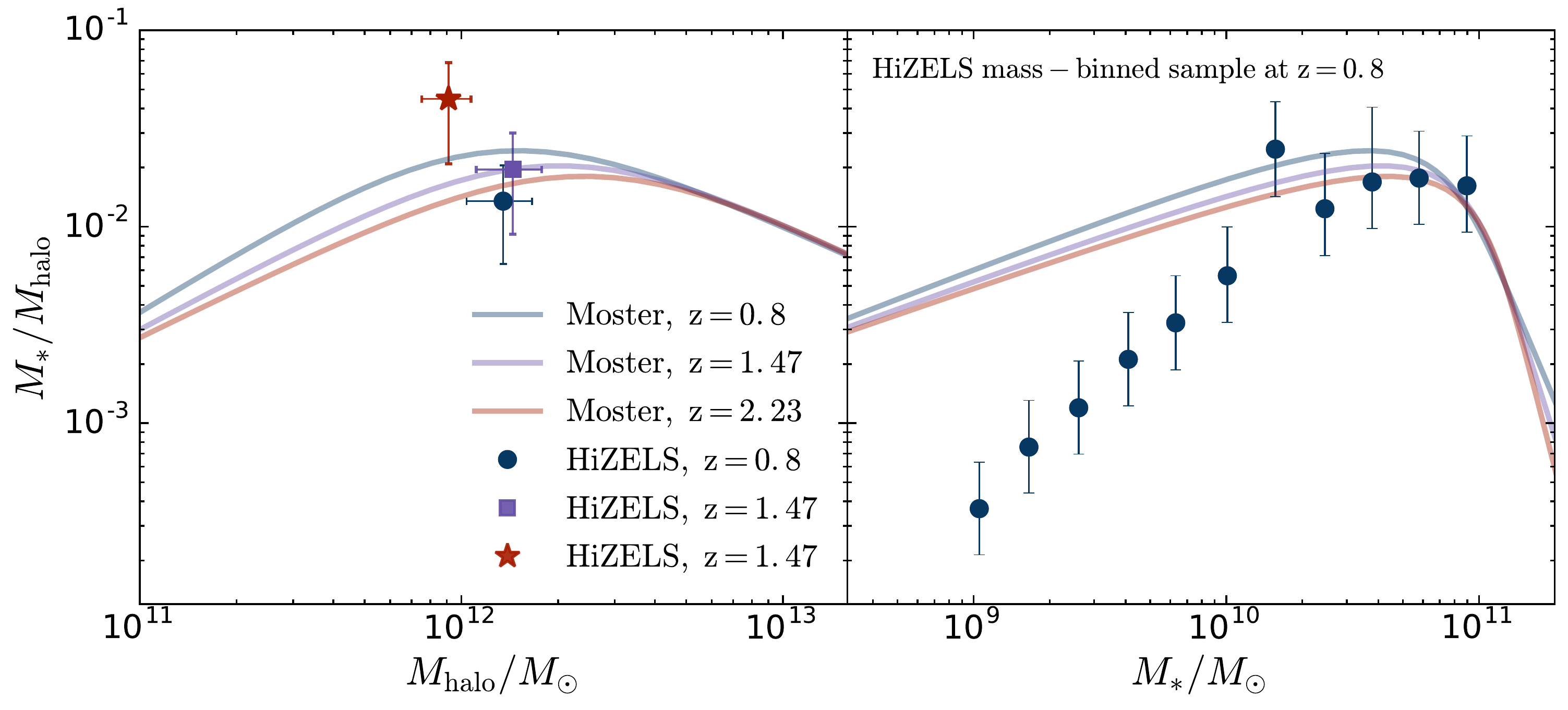}	
	\caption{Left: The stellar mass - halo mass relation from \protect\cite{Moster2013}, with whole HiZELS samples at each redshift overplotted. We use the effective halo mass estimated via the HOD fitting to the whole HiZELS samples at each redshift (see C17). Error bars on the $y$-axis represent the $1\sigma$ uncertainty derived from the MCMC posterior distribution, combined in quadrature with the typical errors on the stellar mass measurements ($0.23,\,0.24$, and $0.26\,\rm{dex}$ for $z=0.8,\,1.47$ and $2.23$ respectively). At all three redshifts, HiZELS galaxies occupy a region at the peak of the SMHR, where conversion of baryons into stellar mass is at a maximum. Right: The stellar mass - halo mass relation from \protect\cite{Moster2013} as a function of stellar mass, with mass-binned HiZELS data from the $z=0.8$ sample within the range $41.72<L_{\rm{H}\alpha}<42.42$ overplotted. While high-mass emitters lie on the relation predicted by \protect\cite{Moster2013}, the lowest mass $\rm{H}\alpha$ emitters lie significantly below it, which indicates that these galaxies are living in more massive haloes than would be expected for central galaxies of their stellar masses.}
	\label{fig:smhmr_hizels}
\end{figure*}
\subsection{The stellar mass-halo mass relation}\label{sec:moster_predictions}
The stellar mass to halo mass ratio (SHMR) is defined as the total stellar mass within a halo divided by the dark matter halo mass. It reflects the relative star formation and satellite galaxy accretion of a halo, compared to its dark matter accretion history, and is effectively a measure of the efficiency of the conversion of baryons into stars. The least massive dark matter haloes build stellar mass inefficiently due to supernova feedback, resulting in low $M_{*}/M_{\rm{halo}}$ fractions. Efficiency appears to increase towards higher halo mass, up to $M_{\rm{halo}}\sim10^{12}M_{\odot}$. A consensus has emerged that haloes of this mass are most efficient at forming stars, with substantial decrease in efficiency above this halo mass \citep[e.g.][]{Behroozi2013,Moster2013}, which is associated with AGN feedback. \cite{Birrer2014} find that the reduced stellar-to-halo mass ratio can be accounted for at high halo masses by the quenching of massive galaxies at around $M^{*}$, the knee of the stellar mass function. There is little evidence for redshift evolution in the peak of the SHMR. Here, we review one approach to modelling the SHMR, and compare our measurements to predictions. \\
\indent \cite{Moster2013} follow \cite{Moster2010} in adopting a double power-law parametrization for the SMHR. The four free parameters are fitted using populations of dark matter haloes and galaxies at redshifts from $z=0$ to $z=4$, specifically dark matter halo populations drawn from the Millennium and Millennium-II Simulations \citep{Springel2005,Boylan-kolchin2009} and galaxy populations from \cite{Li2009} at low redshifts and \cite{Gonzalez-perez2008} and \cite{Santini2012} at high redshifts. At each redshift, \cite{Moster2013} initiate an SMHR with a given set of parameters, and use this to simulate the stellar masses of galaxies within the dark matter haloes they draw from the $N$-body simulation at the same redshift. They then compare the stellar masses of their simulated galaxies to the observed stellar mass function, and assign the modelled SMHR a likelihood. They thus optimize the parameters of the SMHR at each redshift. By including observational errors on high-redshift stellar masses, they are able to derive models that agree well with observed stellar mass functions. \\
\indent \cite{Behroozi2010} show (using another stellar mass-limited approach) that there is little difference between the SHMRs at low halo masses ($M_{\rm{halo}}<10^{12}M_{\odot}$) derived when considering the total stellar mass within the halo or just that of the central galaxy. Given that we argued in C17 that the HiZELS samples are dominated by central galaxies, we use the stellar mass of HiZELS galaxies as a proxy for total stellar mass in the halo. We then compare our estimates of dark matter halo mass for HiZELS galaxies to the predictions of \cite{Moster2013}. We take the same samples of galaxies within large $L_{\rm{H}\alpha}/L_{\rm{H}\alpha}^{*}$ bins at each of the three redshifts, as in C17. We estimate average SED masses as in Section \ref{sec:full_sed}, and use the effective halo masses derived from HOD fitting (see Section \ref{sec:halo_fitting}) to place these samples on to the SHMR. The left-hand panel of Figure \ref{fig:smhmr_hizels} shows that our data are in excellent agreement with the predictions of \cite{Moster2013}. At all three redshifts, HiZELS galaxies occupy a region at the peak of the SMHR. They reside in haloes that are able to support maximum conversion of baryons into stellar mass. \\
\indent Nevertheless, these global averages include galaxies spread over $>2\,\rm{dex}$ in stellar mass, so are not necessarily representative of all HiZELS galaxies. To investigate this, in the right-hand panel of Figure \ref{fig:smhmr_hizels} we place mass-selected subsamples of our $z=0.8$ data on to the same relation. When we calculate the SMHR from the mean stellar mass and derived effective halo mass for each subsample, samples of galaxies with $M_{*}>10^{10}M_{\odot}$ lie approximately on the \cite{Moster2013} relation. However, at low stellar masses, our samples lie significantly below this modelled relation. As discussed in Section \ref{sec:mass_AND_lum}, our low-mass galaxies reside in particularly high-mass haloes for central galaxies of their stellar mass. One possible interpretation of this is that it could be indicative of a substantial amount of stellar mass contained in galaxies that are undetectable by HiZELS within the same halo (i.e. our assumption that the halo's total stellar mass is broadly given by the HiZELS stellar mass is wrong). This points towards some of our low-mass galaxies being satellites. In that case, our low-mass galaxies would be highly star-forming satellites of a (more massive) passive central. However, this would go against the conclusion of the HOD modelling in C17 that the majority of HiZELS galaxies are centrals. Alternatively, we could be picking out starbursting low-mass centrals that will soon gain sufficient stellar mass to place them on to the main SHMR. Given only the current HiZELS observational data, it is difficult to distinguish between these scenarios. We will return to this issue in Section \ref{sec:insights_eagle}, where we compare against the EAGLE simulations.	
\section{Comparing our results to simulations}\label{sec:eagle_comparison}
\subsection{Overview of the EAGLE simulation}
Historically, cosmological hydrodynamical simulations have struggled to reproduce observed properties of galaxy populations simultaneously with the same success as semi-analytic models. Observed statistics of galaxy populations such as stellar mass functions, luminosity functions and the detailed properties of individual galaxies such as sizes, bulge/disc masses and star-formation histories were poorly matched \citep[see][for a review]{Somerville}. This is partly an issue of resolution: to maintain the broadest view of galaxies within the large-scale dark matter structure of the Universe, key processes that determine the detailed evolutionary path of individual galaxies such as star formation and feedback are left unresolved. \\
\indent The latest generation of hydrodynamical simulations has made notable strides by attempting to improve the calibration of sub-grid models to observed properties of galaxy populations. The Virgo Consortium's Evolution and Assembly of GaLaxies and their Environments project, EAGLE, comprises a suite of $\Lambda\rm{CDM}$ simulations based on SPH code \small{GADGET 3} \normalsize \citep{Springel2005}. EAGLE represents a significant improvement on previous hydrodynamical simulations due to its simple implementation of energy feedback from both massive stars and AGN. Subgrid models for these processes are calibrated using two main relations at $z=0.1$: the galaxy stellar mass function, and the galaxy-black hole mass relation. EAGLE's success lies in its reproduction of various other observed relations (e.g. galaxy specific star-formation rate distributions, passive fractions and the Tully-Fisher relation; \citealt{Schaye2015}) that are not explicitly used in the calibration. \cite{Artale2017} also find good agreement between the clustering of blue galaxies in EAGLE and those in the GAMA survey, concluding that these simulated and observed galaxies with similar properties occupy dark matter haloes of similar masses.\\
\indent A number of EAGLE simulations are publicly available \citep{McAlpine2015}. Here, we use version Ref-L100N1504, due to its large volume (box of side length $100\rm{Mpc}$, comoving) and particle number (7 billion). We select galaxies at $z=0.87$, close to the $z=0.8$ HiZELS redshift slice.
\subsection{Halo environments of EAGLE galaxies}\label{sec:gen_eagle}
\begin{figure*}
	\centering
	\includegraphics[scale=0.375]{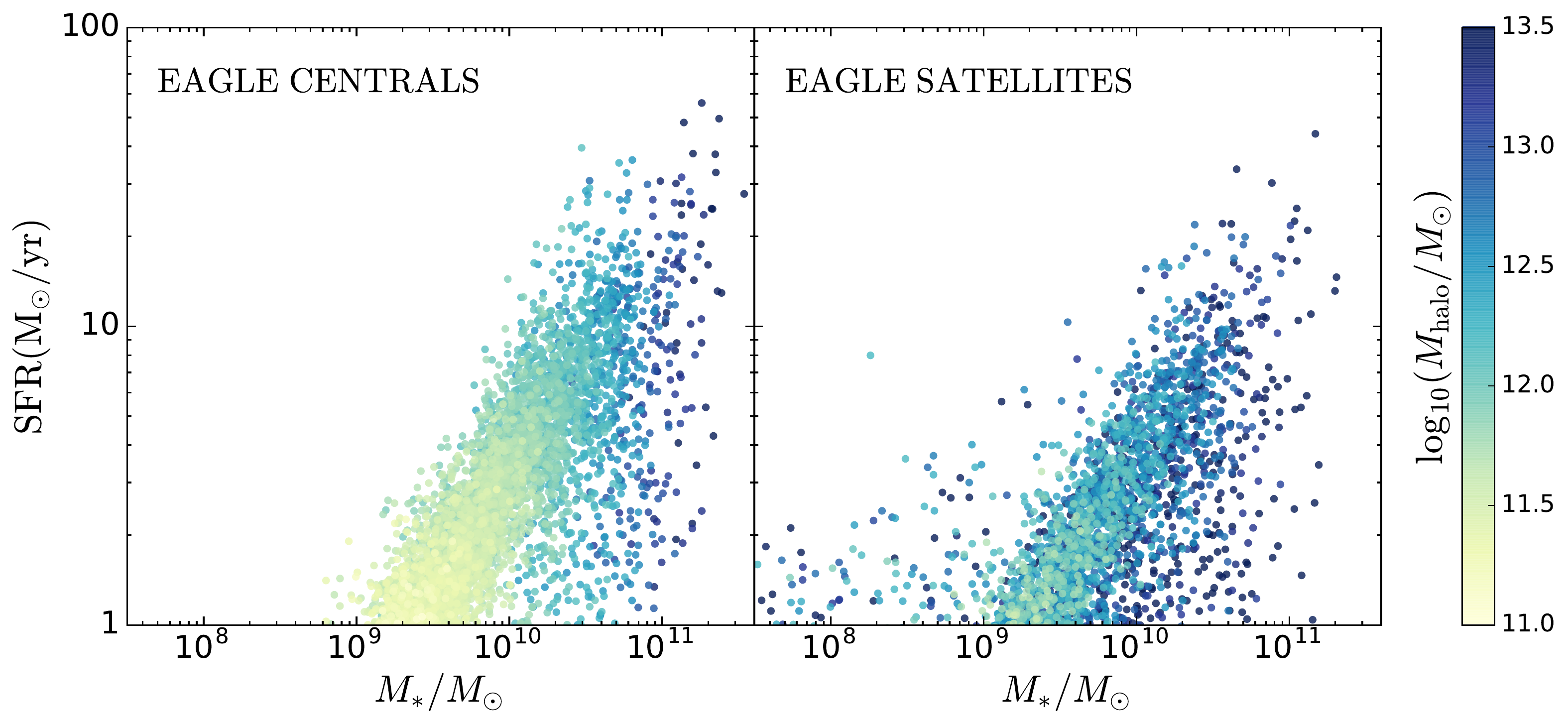}
	\includegraphics[scale=0.375]{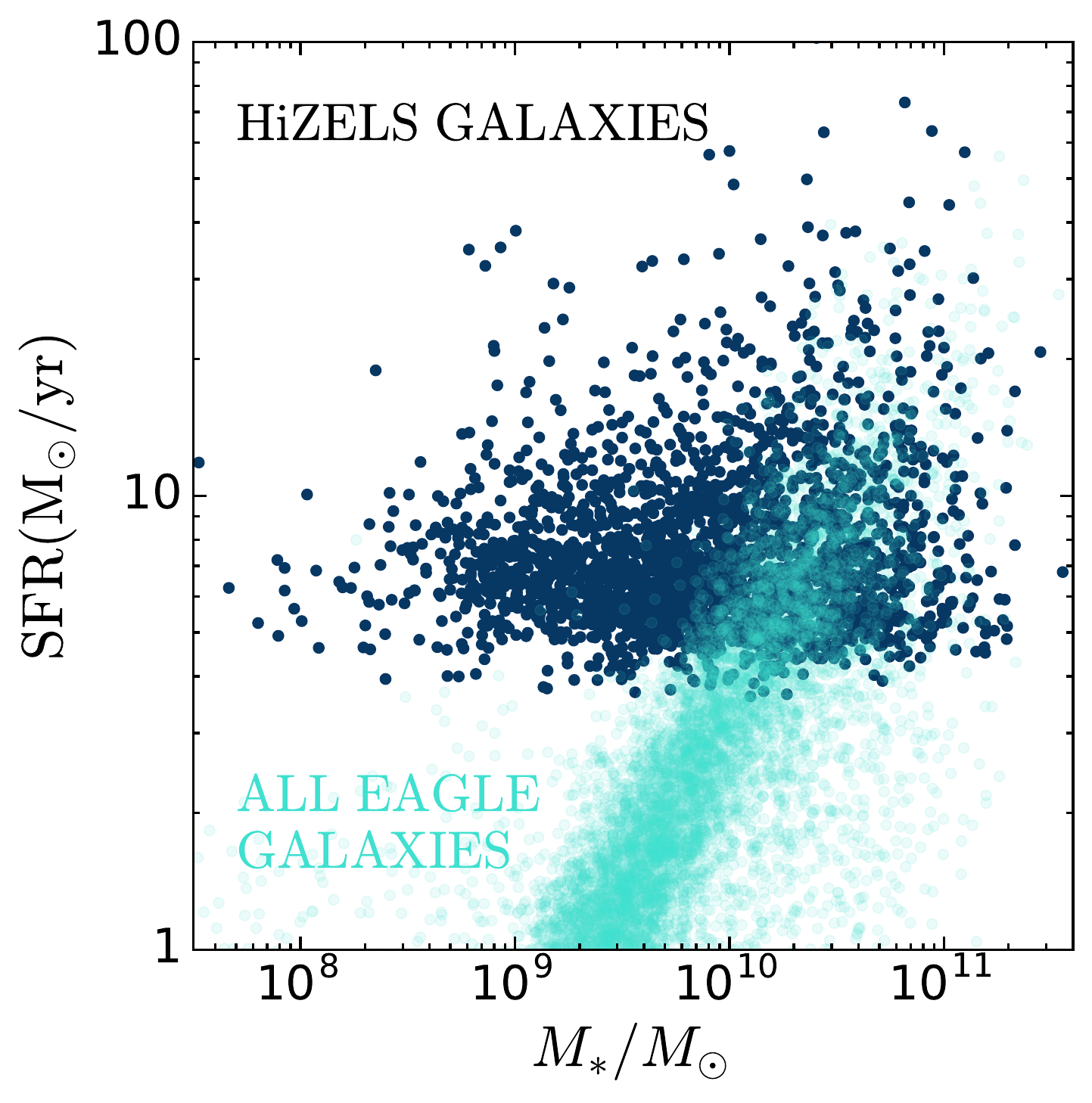}
	\caption{$z=0.87$ galaxies from EAGLE, plotted on the stellar mass - star-formation rate plane using a 30kpc (proper) aperture, colour-coded by their group halo mass. The halo masses of central galaxies (left-hand panel) are strongly correlated with their positions on this plane, with high stellar mass galaxies residing in massive dark matter haloes. The satellite galaxies (middle panel) have greater variance in halo mass at fixed stellar mass, due to the formation of their stellar mass in a smaller halo, before accretion on to more massive haloes. We also show the positions of $z=0.8$ HiZELS galaxies (not colour-coded by halo mass) on the same plane (right-hand panel). HiZELS star-formation rates tend to be slightly higher than those of EAGLE galaxies at low stellar masses.}
    \label{fig:eagle_scatter}
\end{figure*}
\begin{figure} 
	\centering
	\includegraphics[scale=0.54]{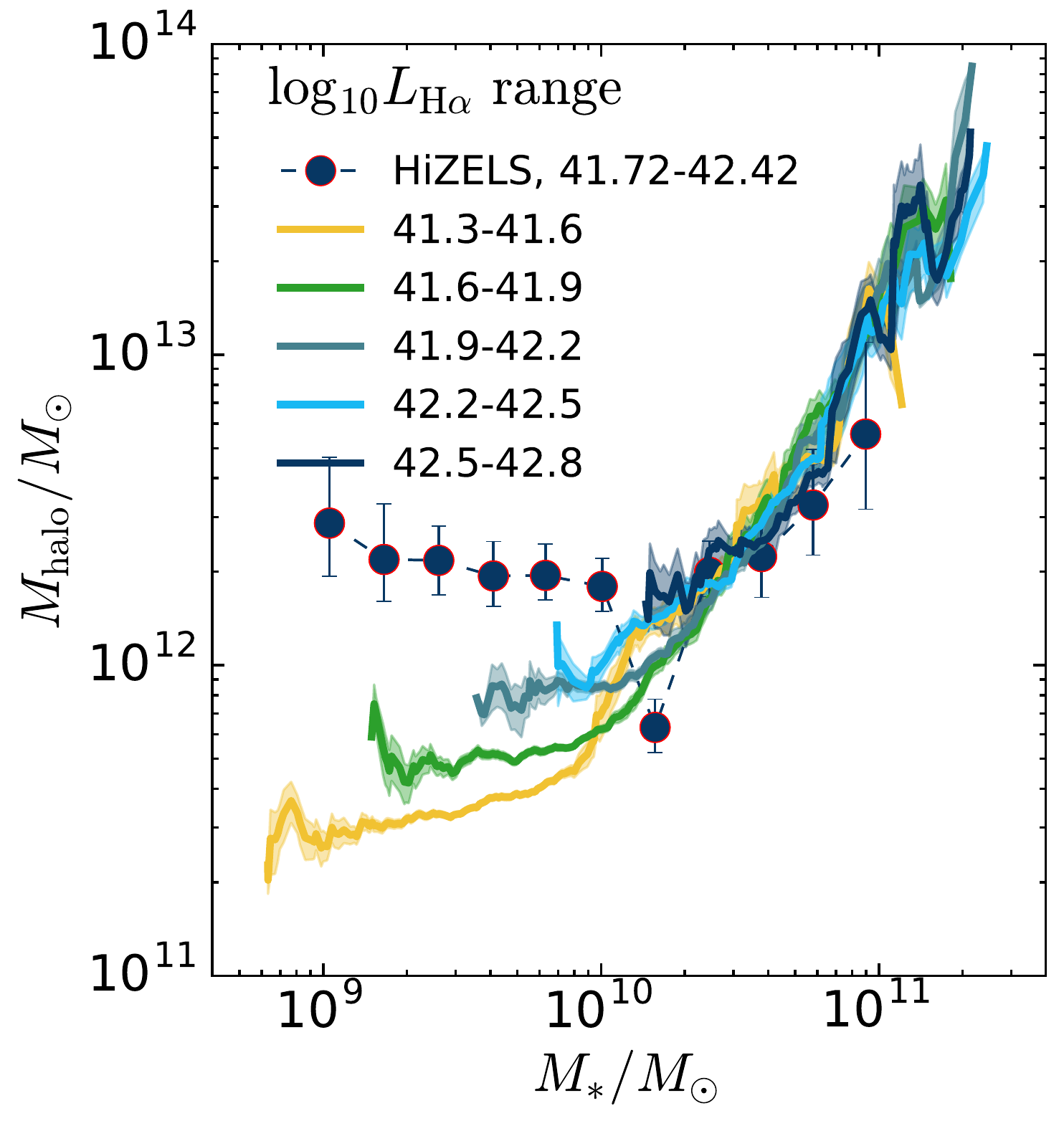}
	\caption{Halo mass as a function of stellar mass for EAGLE central galaxies at $z=0.87$, using moving average bins of size $0.15\,\rm{dex}$. The errors plotted are the standard error on the mean. We select by EAGLE star-formation rate within an aperture of 30kpc (proper), and convert to a rough $L_{\rm{H}\alpha}$ using the \protect\cite{Jr1998} conversion, with correction to a Chabrier IMF. At low stellar masses, the most highly star-forming galaxies lie in more massive haloes than galaxies of the same mass but lower star-formation rates, in line with our HiZELS observations. Low-mass HiZELS galaxies tend to reside in higher mass haloes than even the most highly star-forming EAGLE galaxies. As discussed in Section \protect\ref{sec:gen_eagle}, this could be related to the known $0.2-0.5\,\rm{dex}$ global offset between the EAGLE star-formation rate density and observational measurements.}
    \label{fig:EAGLE_hm_vs_mass}
\end{figure}
Rather than calculating halo mass via the two-point correlation function as we have done for HiZELS galaxies, we identify the halo masses of EAGLE galaxies directly. We use the total friends-of-friends (FOF) mass of the galaxy's halo, labelled as GroupMass in the EAGLE FOF table, as opposed to the subhalo mass. We identify central galaxies as those galaxies for which $\rm{SubGroupNumber}=0$, and satellite galaxies as galaxies with $\rm{SubGroupNumber}>0$. In Figure \ref{fig:eagle_scatter}, we show the typical halo masses of subsamples of EAGLE central and satellite galaxies at $z=0.87$. The stellar mass and star-formation rates used are those within a 30pkpc (proper, as opposed to comoving, kpc) aperture, taken from the EAGLE Aperture table. We see that the halo masses of central galaxies are strongly correlated with their positions on the $\rm{SFR}$-stellar mass plane, with high-stellar mass galaxies residing in massive dark matter haloes. We also see hints of higher halo masses for higher luminosity low-mass central galaxies at fixed stellar mass. We quantify this in more detail in Section \ref{sec:quant_eagle}. For satellite galaxies, halo masses are less strongly correlated with stellar mass or star-formation rate. This reflects the fact that much of a satellite's mass is built up at earlier times, when it is the central of its own subhalo, before this subhalo is accreted on to the larger halo.
\subsection{Mass and star-formation rate dependencies of halo mass from EAGLE}\label{sec:quant_eagle}
In Section \ref{sec:mass_AND_lum}, we showed that at fixed stellar mass, more highly star-forming low-mass galaxies appear more strongly clustered than their less highly star-forming counterparts. Here, we mimic these stellar mass and star-formation rate selections and quantify the average halo masses of EAGLE central galaxies binned in the same way. We convert EAGLE star-formation rates to rough $\rm{H}\alpha$ luminosities, for comparison with HiZELS, using the \cite{Jr1998} $L_{\rm{H}\alpha}-\rm{SFR}$ conversion given in Section \ref{sec:ha_sample} and assuming the same \citet{Chabrier2003} IMF as used by EAGLE. \\
\indent Our results are presented in Figure \ref{fig:EAGLE_hm_vs_mass}. We see a strong $M_{*}-M_{\rm{halo}}$ correlation at high stellar masses, which flattens at low stellar masses, just like we found for the HiZELS samples. At low stellar masses ($M_{*} \lesssim 10^{10}M_{\odot}$), average halo mass increases with star-formation rate at fixed stellar mass. At high stellar masses ($M_{*} \gtrsim 10^{10}M_{\odot}$), average halo mass is roughly independent of star-formation rate for central galaxies. This is broadly consistent with our HiZELS observational results. However, there appears to be a lack of very highly star-forming, low-mass galaxies in EAGLE (cf. Figure \ref{fig:eagle_scatter}). EAGLE galaxies do not reach the high luminosities of HiZELS galaxies, perhaps because of insufficiently bursty star formation in the simulations, or the inability to resolve bursts on small time-scales. There are well-known tensions between EAGLE star-formation rates and observations. The specific star-formation rates of EAGLE star-forming galaxies are $0.2-0.5\,\rm{dex}$ below those inferred from observations, across all redshifts \citep{Furlong2015a}. Despite the offset in global star-formation rate density, applying the required $0.3\,\rm{dex}$ star-formation rate offset to all star-formation rates would break the agreement between simulated and observed stellar mass densities. Nevertheless, the broad trends of our observational results are supported by EAGLE: for low stellar mass central galaxies, galaxy dark matter halo mass is not a simple function of stellar mass, but also depends on the galaxy's star-formation rate.  
\subsection{Physical interpretation using EAGLE}
\begin{figure*} 
	\centering
	\includegraphics[scale=0.4]{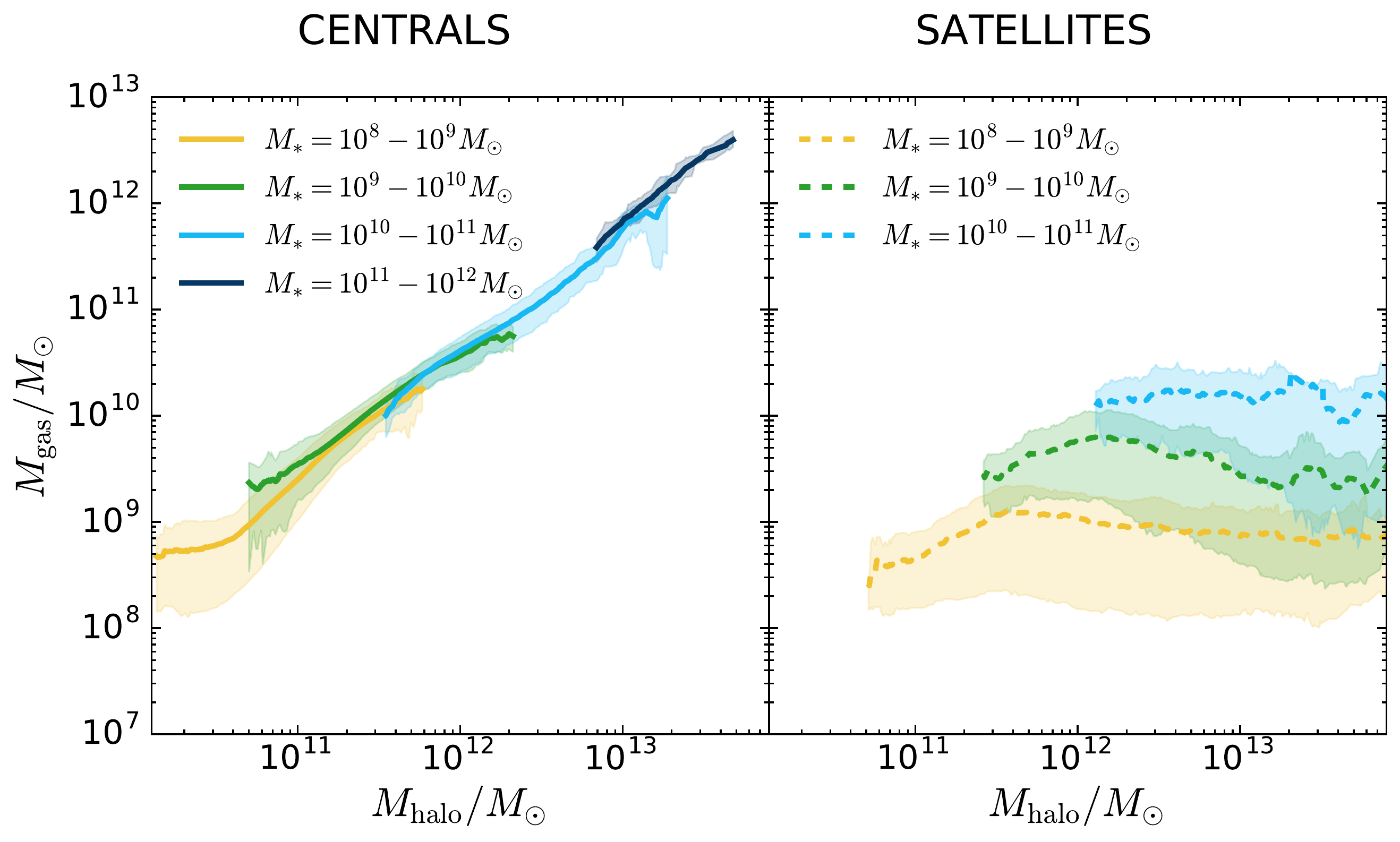}
	\includegraphics[scale=0.4]{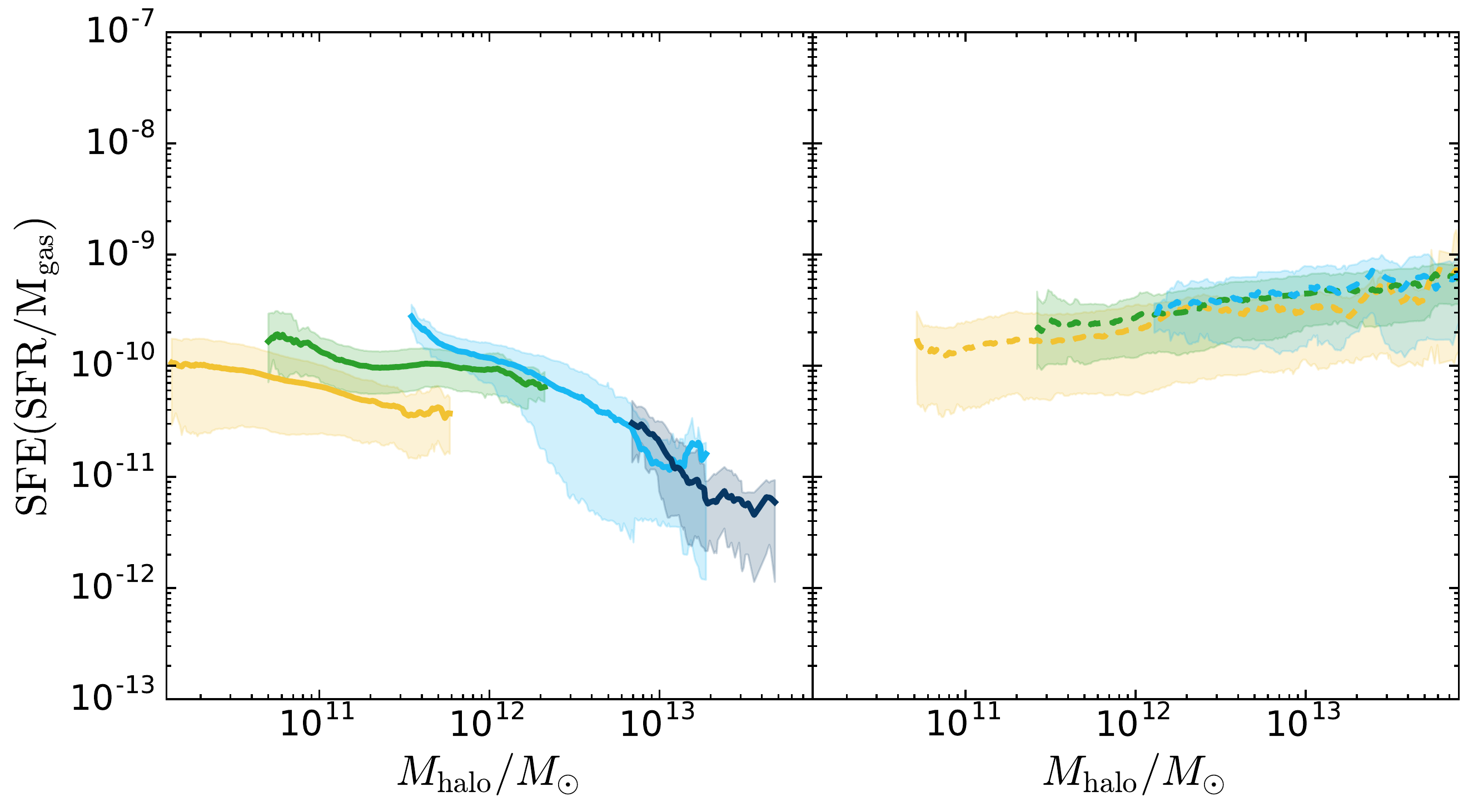}
	\includegraphics[scale=0.4]{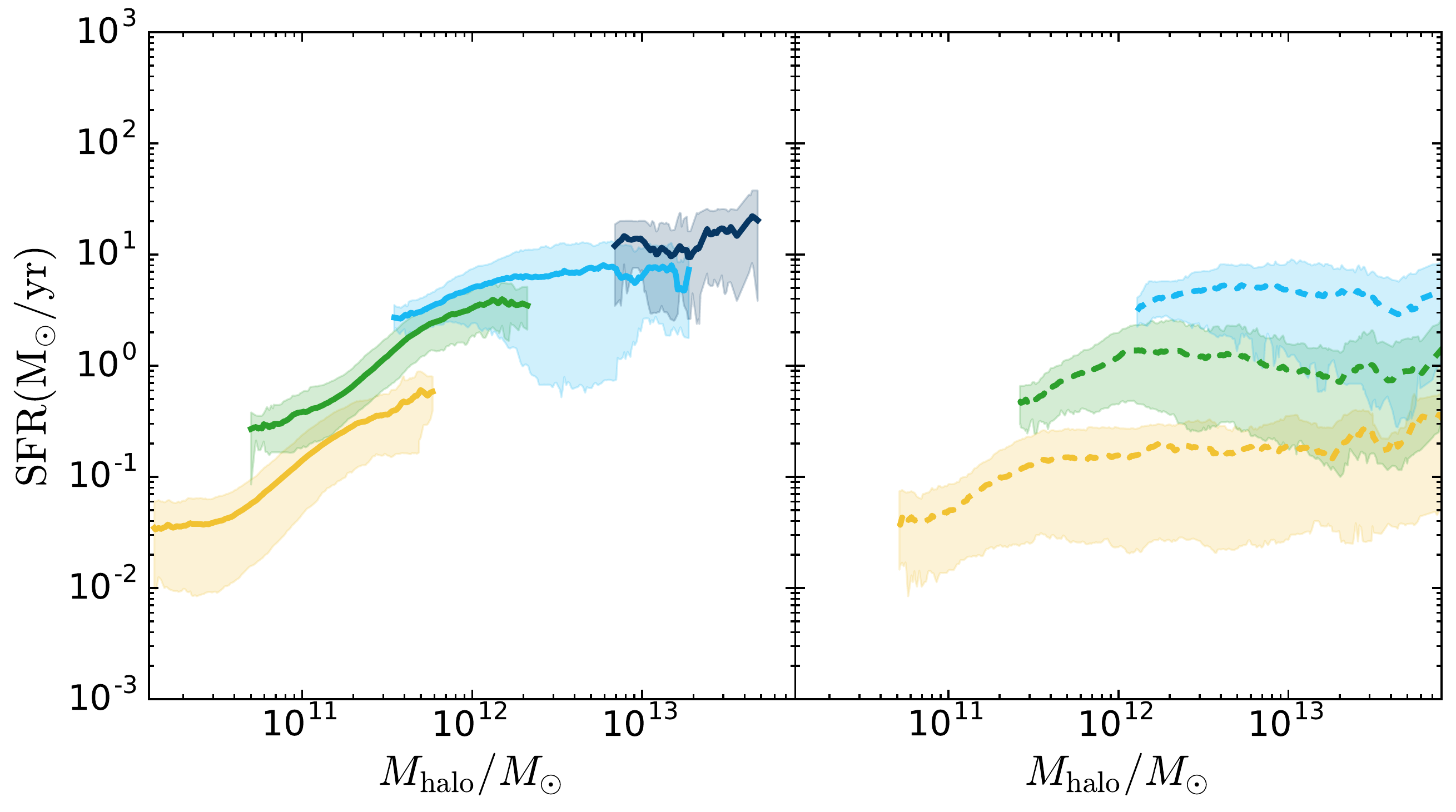}
	\caption{Mean gas mass, star-formation efficiency and star-formation rate as a function of halo mass for satellite and central EAGLE galaxies at $z=0.8$, with $1\sigma$ error contours. For central galaxies at all stellar masses, galaxy gas mass correlates tightly with host halo mass. Although star-formation efficiency decreases with increasing halo mass, mean star-formation rate increases with halo mass, for central galaxies in haloes with $M_{\rm{halo}}<10^{12}M_{\odot}$. Dependencies on stellar mass are weak by comparison. In contrast, for satellites, star-formation rate does not depend strongly on $M_{\rm{halo}}$, but more on $M_{*}$.}
    \label{fig:interp_plots}
\end{figure*}	
Here, we use EAGLE to investigate why our most highly star-forming HiZELS galaxies tend to reside in the most massive dark matter haloes. We study the average gas content, $M_{\rm{gas}}$, star-formation rate, $\rm{SFR}$, and star-formation efficiency, $\rm{SFE} = \frac{\rm{SFR}}{M_{\rm{gas}}}$ (the inverse of the gas depletion time-scale), as a function of halo mass and stellar mass. We include only galaxies with $\rm{SFR}>0$ in this analysis. Figure \ref{fig:interp_plots} shows our results. The $\log_{10}M_{\rm{halo}}-\log_{10}M_{\rm{gas}}$ relation for central galaxies is linear, and independent of galaxy stellar mass. At all stellar masses, the most massive haloes supply the most gas to their centrals. The same relation is strikingly different for satellite galaxies: the average gas mass of a satellite galaxy appears broadly independent of its halo mass, but varies significantly with stellar mass. At fixed halo mass, more massive satellite galaxies have larger gas reservoirs. This is likely due to the gas content being established earlier, prior to accretion on to a more massive halo, when the satellite galaxy's gas mass would have correlated with the mass of its subhalo (using the mass of the EAGLE subhalo places centrals and satellites on to the same sequence), which in turn correlates more closely with stellar mass. \cite{Wetzel2013} argue that satellite galaxies retain their cold gas reservoirs upon infall and continue to form stars on long time-scales. This is broadly supported by EAGLE, where the gas mass of satellites of fixed stellar mass varies little with halo mass. The role of gas stripping in these galaxies' evolution appears to be sub-dominant. \\
\indent The star-formation efficiencies of central and satellite galaxies are also markedly different. $\rm{SFE}$ falls with increasing halo mass for central galaxies at all stellar masses, with a particularly steep decrease above $M_{\rm{halo}}\sim10^{12}M_{\odot}$. Higher stellar mass centrals also have slightly higher star-formation efficiencies, particularly in the lowest mass haloes. Satellite galaxies display a weak increase in $\rm{SFE}$ with halo mass ($\sim1\,\rm{dex}$ over $\sim3\,\rm{dex}$ in $M_{\rm{halo}}$), independently of stellar mass, perhaps due to increased intracluster medium pressure in higher mass haloes (e.g. \citealt{Bekki2014}). \\
\indent The bottom row of Figure \ref{fig:interp_plots} shows the combination of the gas content and star-formation efficiency: the mean star-formation rate as a function of halo mass. Below $M_{\rm{halo}} \sim 10^{12}M_{\odot}$, mean $\rm{SFR}$ increases with $M_{\rm{halo}}$ for central galaxies of all stellar masses. This increase appears to be driven by gas content: gas cooling from the halo fuels star formation in central galaxies, with higher cooling rates in more massive haloes and little variation in star-formation efficiency. At fixed halo mass, the more massive galaxies have higher $\rm{SFRs}$ due to increasing efficiency of gas conversion. Above $M_{\rm{halo}} \sim 10^{12}M_{\odot}$, the $\rm{SFR}-M_{\rm{halo}}$ relation appears to flatten due to decreasing star-formation efficiency; there are also few star-forming galaxies at these high halo masses. Satellite galaxies display a very weak increase in $\rm{SFR}$ with halo mass at the lowest halo masses, and subsequent flattening at high halo masses. This appears to be driven by a combination of increasing star-formation efficiency and decreasing gas content with increasing halo mass. At fixed halo mass, more massive satellites are more highly star-forming due to their higher gas content. \\
\indent EAGLE thus provides insights into the drivers of the trends we observe with HiZELS. Simulated low-mass, highly star-forming galaxies also reside in higher mass haloes than their less highly star-forming counterparts. EAGLE shows that these trends are likely driven by gas supply rather than increased star-formation efficiencies in high-mass haloes. One remaining tension is the paucity of very highly star-forming galaxies in EAGLE compared to those observed. Those EAGLE galaxies that are highly star-forming tend to be satellites (see Figure \ref{fig:eagle_scatter}). Given the difficulties in an auto-correlation analysis of distinguishing star-forming satellites of passive centrals from star-forming centrals given only a star-formation rate-selected sample, there are significant uncertainties in our satellite fraction determination discussed in C17. Nevertheless, the scarcity of highly star-forming centrals in EAGLE may well be due to star formation in the high redshift Universe being more bursty and stochastic than is simulated or recorded in the timestep-smoothed EAGLE output. 
\subsection{Insights into the SHMR from EAGLE}\label{sec:insights_eagle}
\begin{figure}
	\centering
	\includegraphics[scale=0.55]{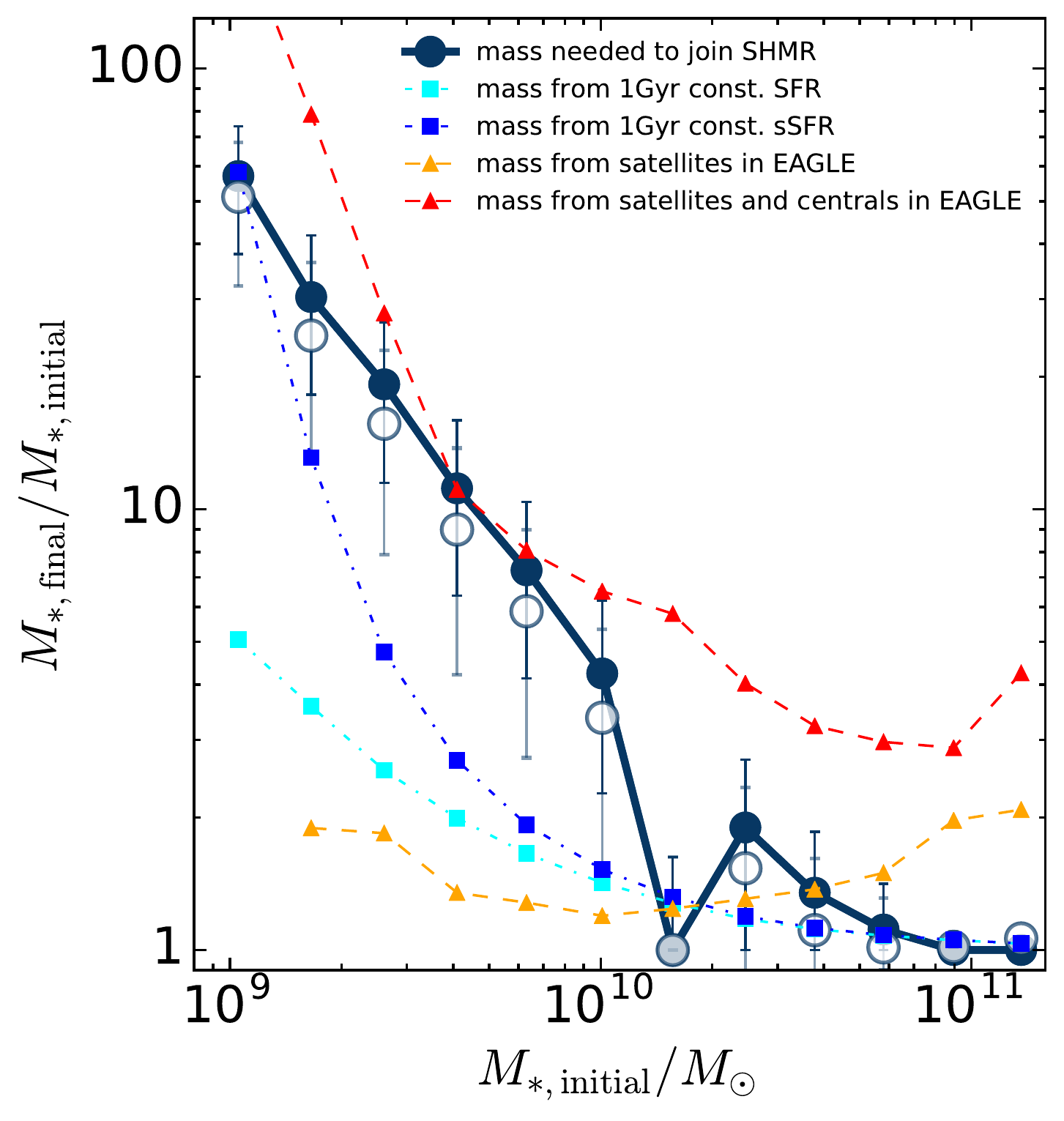}
	\caption{The growth factor required, as a function of stellar mass, to bring the $z=0.8$ HiZELS galaxies on to the SHMR (thick blue line). Closed circles use the SHMR relation from \protect\cite{Moster2013}, and open circles use the SHMR constructed using EAGLE. This indicates the approximate uncertainty on the SHMR itself. We model corrections to the mass of the HiZELS galaxy obtained under the assumption of $1\rm{Gyr}$ star formation at the measured star-formation rate and specific star-formation rate. For comparison, the other lines show the simulated corrections to the mass contained in the dark matter host haloes of HiZELS galaxies using EAGLE. High-mass HiZELS galaxies already lie on the SHMR. Low-mass EAGLE galaxies ($M_{*}<10^{10}M_{\odot}$) with comparable star-formation rates reside in dark matter haloes with significant stellar mass contributions from companion galaxies. A correction from these places HiZELS galaxies on or above the main SMHR.}
	\label{fig:shmhr_corrs_sim}
\end{figure}
In Section \ref{sec:moster_predictions}, we placed our HiZELS samples on to the SHMR, considering the typical halo mass derived from clustering measurements for galaxies in different stellar mass bins. We found that mass-selected subsamples of HiZELS galaxies tend to lie below the SHMR at the lowest stellar masses. We suggested that this could be due to significant additional stellar mass within the same haloes, indicating that some of our low-mass galaxies are satellites of central galaxies which lie below the HiZELS $\rm{H}\alpha$ detection limits. Alternatively, these galaxies could be very highly star-forming centrals which will soon gain enough mass to place them on to the main SHMR. Here, we investigate these scenarios, to ascertain whether either star formation at HiZELS observed rates or unaccounted stellar mass within the same halo (as estimated using the EAGLE simulations) can account for the additional stellar mass needed.\\
\indent We begin by calculating the increase in stellar mass required to move our HiZELS measurements diagonally on to the \cite{Moster2013} SHMR, assuming little change in halo mass. For moderate to high-mass galaxies ($M_{*}=10^{10}-10^{11}M_{\odot}$, the SHMR offsets are very small, but we find higher offsets (factors of tens) for galaxies at lower stellar masses. The required growth factors are shown as a function of stellar mass in Figure \ref{fig:shmhr_corrs_sim}. \\
\indent Next, we use the average $L_{\rm{H}\alpha}$ within each stellar mass bin to calculate a typical stellar mass increase over $1\rm{Gyr}$ of star formation if either the current star-formation rate or the current specific star-formation rate is maintained.\\
\indent Finally, we select a sample of galaxies in EAGLE with comparable $\rm{SFRs}$ to those observed by HiZELS to evaluate the mass contribution of other galaxies in the halo. We do this in two ways. The first selects only star-forming central galaxies. This is motivated by C17, which estimated low satellite fractions for these samples. The second allows our star-forming EAGLE comparison galaxies to be either centrals or satellites. For each EAGLE comparison sample, we identify other EAGLE galaxies within the same dark matter haloes, and calculate a stellar mass correction, the difference between the stellar mass in the detected star-forming galaxy and the total stellar mass in the halo. These correction factors are shown in Figure \ref{fig:shmhr_corrs_sim}.\\
\indent Figure \ref{fig:shmhr_corrs_sim} shows that for the high-mass galaxies, which already lie on the SHMR, stellar mass is little affected by $\sim1\rm{Gyr}$ of star formation at either fixed SFR or fixed sSFR, and that similarly accounting for satellite galaxies makes little difference to the stellar mass of the haloes. At lower stellar masses, ongoing star formation at fixed SFR over $\sim1\rm{Gyr}$ time scales can produce a significant increase in stellar mass (up to a factor of a few), but falls far short of that required to bring the galaxies on to the SHMR. Likewise, $1\rm{Gyr}$ of star formation at fixed sSFR or considering the contribution of satellite galaxies in the same halo, both appear insufficient. Instead, it appears likely that some contribution from centrals within the same halo is required if our samples are going to move on to the SHMR, indicating that a proportion of our low-mass star-forming galaxies may be satellites of centrals with lower SFRs. Otherwise, we are detecting low-mass central galaxies that lie significantly below the SHMR, and will remain so for more than a $\rm{Gyr}$, even if they maintain their current high specific star-formation rates.
\section{Conclusions}\label{sec:conclusions}
We have studied the clustering of intermediate redshift star-forming galaxies and its dependence on star-formation rate and stellar mass. Our samples comprise $\rm{H}\alpha$-selected galaxies predominantly on and above the star-forming main sequence at three redshifts, $z = 0.8,\,1.47$ and $2.23$. We summarize the key results here. 
\begin{itemize}
\item  At all three redshifts, we find clear evidence for a monotonic increase in clustering strength, $r_{0}$, with stellar mass above $M_{*}\sim2-3\times10^{10}M_{\odot}$. At lower stellar masses, where star-forming galaxies selected by HiZELS lie significantly above the main sequence, this relation flattens. The $M_{*}-r_{0}$ relation is very different from the $\log_{10}L_{\rm{H}\alpha}-r_{0}$ relation studied in C17, which shows a significant and monotonic increase of $r_{0}$ with increasing $\rm{H}\alpha$ luminosity, with no flattening at the lowest luminosities. \\
\item At fixed stellar mass, higher $\rm{H}\alpha$ luminosity subsamples are more strongly clustered than their less luminous counterparts. This is particularly pronounced at the lowest stellar masses ($M_{*} < 10^{10}M_{\odot}$). We find consistent results when we mimic our $L_{\rm{H}\alpha}$ cuts using the EAGLE simulations. We deduce that these highly star-forming low-mass galaxies are undergoing environmentally driven star formation. Investigating the cause of this using EAGLE reveals that our trends are likely driven by enhanced gas supply in small groups compared to the field. \\
\item  We compare our mass-binned clustering measurements of $L_{\rm{H}\alpha}$-selected galaxies to those obtained from mass-selected samples, and show that measurements of galaxy clustering are strongly dependent on the galaxy selection criteria. We find that HiZELS star-forming galaxies are less strongly clustered than mass-selected galaxies at fixed stellar mass. Compilations of literature measurements confirm that passive and mass-selected samples tend to be more strongly clustered than star-forming samples back to at least $z\sim2$. Mass-selected samples seem to be picking up many more quenched satellites in massive haloes. We argue that our results are in line with average star-formation rates increasing towards group densities but decreasing at the highest cluster densities, where environmentally driven quenching plays a stronger role. \\
\item We place HiZELS samples on the SHMR obtained empirically using mass-selected galaxy samples by \cite{Moster2013}. We find that, on average, these highly star-forming galaxies lie at its peak, where baryon to stellar mass conversion is most efficient. Extending this to mass-binned subsamples, we show that high-mass HiZELS galaxies ($M_{*}>10^{10}M_{\odot}$) lie on the SHMR, but that at lower stellar masses, our samples lie below the relation.\\
\item Finally, we consider the effect of ongoing star formation and show that current star-formation rates are insufficient to return low-mass galaxies to the SHMR. Using EAGLE, we find that if a proportion of these are satellites, typical stellar mass corrections from HiZELS-undetected galaxies within the same haloes can easily bring low-mass galaxies up on to the main SHMR. 
\end{itemize}
In conclusion, we use the clustering of carefully selected star-forming galaxies with well-defined redshift distributions to determine their typical halo masses. We present evidence for environmentally driven star formation in low-mass galaxies, some of which lie well above the main sequence. We use the EAGLE simulation to strengthen the physical interpretation, and show that it is likely that these star-formation rates are driven by increased gas content in galaxies residing in higher mass haloes. 
\section*{Acknowledgements}
This work is based on observations obtained using the Wide Field CAMera (WFCAM) on the 3.8-m United Kingdom Infrared Telescope (UKIRT), as part of the High-redshift(Z) Emission Line Survey (HiZELS; U/CMP/3 and U/10B/07). It also relies on observations conducted with HAWK-I on the ESO Very Large Telescope (VLT), programme 086.7878.A, and observations obtained with Suprime-Cam on the Subaru Telescope (S10B-144S). We acknowledge the Virgo Consortium for making their simulation data available. The EAGLE simulations were performed using the DiRAC-2 facility at Durham, managed by the ICC, and the PRACE facility Curie based in France at TGCC, CEA, Bruy\`{e}res-le-Ch\^{a}tel. \\
\indent We thank the anonymous reviewer for their detailed suggestions. We are grateful to Steven Murray for making the HALOMOD and HMF {\small{PYTHON}} packages available and for guidance on their use. We also thank Stuart McAlpine for help with the EAGLE simulations.\\
\indent RKC acknowledges funding from an STFC studentship. PNB is grateful for support from STFC via grant ST/M001229/1. DS acknowledges financial support from the Netherlands Organisation for Scientific research (NWO) through a Veni fellowship and from Lancaster University through an Early Career Internal Grant A100679. IRS acknowledges support from STFC (ST/P000541/1), the ERC Advanced Investigator programme DUSTYGAL 321334, and a Royal Society/Wolfson Merit Award. 

\bibliographystyle{mnras}
\bibliography{Edinburgh}


\bsp	
\label{lastpage}
\end{document}